\documentclass[journal]{IEEEtran}
\usepackage{graphicx,epsfig,amsmath,amssymb,bm}

\def\LSB{\left[}        
\def\RSB{\right]}       
\def\LB{\left(}         
\def\RB{\right)}        

\newfont{\bbb}{msbm10 scaled 500}

\newcommand{\av}{{\bf a}}
\newcommand{\bv}{{\bf b}}

\newcommand{\ev}{{\bf e}}

\newcommand{\iv}{{\bf i}}

\newcommand{\rv}{{\bf r}}

\newcommand{\uv}{{\bf u}}
\newcommand{\wv}{{\bf w}}
\newcommand{\vv}{{\bf v}}
\newcommand{\xv}{{\bf x}}
\newcommand{\yv}{{\bf y}}

\newcommand{\Am}{{\bf A}}
\newcommand{\Bm}{{\bf B}}
\newcommand{\Cm}{{\bf C}}

\newcommand{\Id}{{\bf I}}

\newcommand{\Km}{{\bf K}}

\newcommand{\Pm}{{\bf P}}
\newcommand{\Qm}{{\bf Q}}
\newcommand{\Rm}{{\bf R}}

\newcommand{\Wm}{{\bf W}}

\newcommand{\Xm}{{\bf X}}
\newcommand{\Ym}{{\bf Y}}

\newcommand{\deltav}{\hbox{\boldmath$\delta$}}

\newcommand{\phiv}{\hbox{\boldmath$\phi$}}

\newcommand{\thetav}{\hbox{\boldmath$\theta$}}

\newcommand{\Sigmam}{\hbox{\boldmath$\Sigma$}}

\usepackage{subfigure}

\newcommand{\beqa}{\begin{eqnarray}}
\newcommand{\eeqa}{\end{eqnarray}}
\newcommand{\dsp}{\displaystyle}
\usepackage{bbm}
\usepackage[ruled]{algorithm2e}

\def\argmax{\operatornamewithlimits{arg\,max}}

\begin{document}

\title{Performance and Resilience of Cyber-Physical Control Systems with
Reactive Attack Mitigation}
\author{Subhash~Lakshminarayana~\IEEEmembership{Member, IEEE}, Jabir Shabbir Karachiwala, Teo Zhan Teng,  Rui Tan~\IEEEmembership{Senior Member, IEEE}, and David K.Y. Yau~\IEEEmembership{Senior Member, IEEE}
\thanks{S. Lakshminarayana is with the University of Warwick, UK (email: subhash.lakshminarayana@warwick.ac.uk). J.S. Karachiwala is with Oracle India Pvt. Ltd. (email: jabirrokx@yahoo.com). Z. Teng is with GovTech Singapore (email: teozt@hotmail.com). R. Tan is with the Nanyang Technological University, Singapore (email: tanrui@ntu.edu.sg). D.K.Y. Yau is with the Singapore University of Technology and Design, Singapore (email: david\_yau@sutd.edu.sg). 
\newline This research was supported in part by the National Research Foundation (NRF),
Prime Minister’s Office, Singapore, under its National Cybersecurity R\&D
Programme (Award No. NRF2014NCR-NCR001-31), in part by the the Energy Programme and administered by the Energy Market Authority (EP award no. NRF2017EWT-EP003-06), in part by SUTD-ZJU IDEA award no. (VP) 201805, and in part by a start-up grant at NTU. We acknowledge the support of NVIDIA Corporation with the donation of two GPUs used in this research. We also acknowledge Zhenyu Yan for managing the computation resources used in this paper.
\newline The paper was presented in part at the ACM Int'l Conf. Future Energy Systems (e-Energy), 2017 \cite{LaksheEnergy2017}.
}\vspace{-2em}}
\maketitle	
\begin{abstract}
This paper studies the performance and resilience of a linear cyber-physical control system (CPCS) with attack detection and reactive attack mitigation in the context of power grids. It addresses the problem of deriving an optimal sequence of false data injection attacks that maximizes the state estimation error of the power system. The results provide basic understanding about the limit of the attack impact. The design of the optimal attack is based on a Markov decision process (MDP) formulation, which is solved efficiently using the value iteration method. We apply the proposed framework to the voltage control system of power grids and run extensive simulations using PowerWorld. The results show that our framework can accurately characterize the maximum state estimation errors caused by an attacker who carefully designs the attack sequence to strike a balance between the attack magnitude and stealthiness, due to the simultaneous presence of attack detection and mitigation. Moreover, based on the proposed framework, we analyze the impact of false positives and negatives in detecting attacks on the system performance. The results are important for the system defenders in the joint design of attack detection and mitigation to reduce the impact of these attack detection errors.Finally, as MDP solutions are not scalable for high-dimensional systems, we apply Q-learning with linear and non-linear (neural networks based) function approximators to solve the attacker's problem in these systems and compare their performances.
\end{abstract}


\section{Introduction}
Critical infrastructures such as power grids are increasingly becoming the target of cyber attacks as evidenced in recent high-profile incidents, e.g., the BlackEnergy
attack \cite{Ukraine2016}. These attacks injected false sensor data and/or control commands to the industrial control systems and resulted in widespread damage to the physical infrastructures and service outages.  These incidents alert us to a general class of attacks called {\em false data injection} (FDI) against cyber-physical systems (CPS).

Attack detection and mitigation are two basic CPS security research problems, where the {\em attack detection} makes decisions in real time regarding the presence of an attack and {\em attack mitigation} isolates a detected attack and/or reduces its adverse impact on the system performance. 
CPSs often have various built-in anomaly detection methods that are effective in detecting simple fault-like FDI attacks, such as injecting surges, ramps, and random noises. However, critical CPSs (e.g., power grids)
are the target of sophisticated attackers (such as hostile national organizations), 
whose attacks are often well-crafted using detailed knowledge of the system and its anomaly detection methods. To avoid detection, the attacker can inject a sequence of attacks of small magnitude and gradually mislead the system to a sub-optimal and even unsafe state. 
However, due to the stochastic nature of the physical and measurement processes of CPSs, as well as the adoption of stringent, advanced attack detectors, the well-crafted attacks can be detected probabilistically \cite{Mo2015}. Upon the detection, mitigation should be activated to isolate the attack or maintain acceptable system performance in coexisting with the attack. 

Therefore, attack detection and mitigation are deeply coupled and they jointly define the system resilience against FDI attacks. On the one hand, a conservative detector may miss attacks, causing system performance degradation due to the mis-activation of attack mitigation. On the other hand, an aggressive detector may frequently raise false positives, triggering unnecessary mitigation actions in the absence of attacks, while attack mitigation generally needs to sacrifice the system performance to increase its robustness against attacks. Thus, it is important to understand the joint effect of attack detection and mitigation on the system performance, which serves as a basis for designing satisfactory detection-mitigation mechanisms. 
However, prior research on FDI attacks mostly study attack detection and mitigation separately \cite{Kwon2013,Mo2015,BaiGupta2014}, and falls short of capturing their joint effect on the system. The studies on attack detection \cite{Kwon2013,Mo2015} generally ignore the attack mitigation triggered by probabilistic detection of attacks, and its impact on the future system states. On the other hand, the studies on attack mitigation \cite{Barreto2013,ZhuBasar2015} assume that the attack has been detected, and ignore the probabilistic nature of the attack detection and any adverse impact of mis-activation or false activation of mitigation due to missed detections (MDs) and false positives (FPs).

As an early (but important) effort in closing the gap, we jointly consider attack detection and mitigation in the system defense and study their joint effect from both the attacker and the defender's perspectives. In particular, from the attacker's perspective, we  investigate the largest system performance degradation that a sophisticated attacker can cause in the presence of such a detection-mitigation defense mechanism. Studying this largest performance degradation helps us quantify the limit of attack impact. From the defender's perspective, we quantify the system performance degradation due to FPs and MDs. The framework serves as an important basis for designing and assessing attack detection and mitigation strategies for critical infrastructures.

Although our analysis applies to general control systems, the primary focus of this work is power grid controls. We consider a general discrete-time linear time invariant (LTI) system with a feedback controller that computes its control decision based on the system state estimated by a Kalman filter (KF). The LTI model is a widely used approximation to model many control loops in a power grid, such as the power grid's voltage control system \cite{Paul1987, Ilic1995} and the generator swing equations \cite{Kundur1994}. Moreover, under the assumption of direct current (DC) power flows, the observation mode can also be linearized \cite{wood1996power, MonticelliStEst2000}. These LTI models have been widely used in power grid research literature \cite{FabioPow2011, Zhao2014, Amini2018}.
For each time step, the controller uses a $\chi^2$ attack detector \cite{MEHRAChiSquare1971}, and activates mitigation actions upon detecting an attack. 
Following the Kerckhoffs's principle, we consider an attacker who accurately knows the system and its attack detection and mitigation methods. The attacker launches FDI attacks on the sensor measurements over an attack time horizon, aiming at misleading the controller into making erroneous control decisions. 

As the attack detection at each time step is probabilistic, we formulate the attacker's problem as a constrained stochastic optimization problem with an objective of maximizing the state estimation error over the attack time horizon, subject to a general constraint that the energy of the attack signal is upper-bounded. 
The attacker faces a fundamental dilemma in designing his attack -- a large attack magnitude will result in high detection probability, thus nullifying the attack impact on the system (due to mitigation) whereas a small attack magnitude increases stealthiness but may cause little damage. Thus, the solution to this problem leads to an attack sequence that strikes a balance between attack magnitude and stealthiness to achieve the largest system performance degradation.

The main challenge in solving the aforementioned attacker's problem lies in the fact that the system state at any time depends on all the past attack detection results, due to reactive attack mitigation. Thus, the optimal attack at any time must exhaustively account for all possible sequences of past detection results, which is computationally complex. Moreover, the probabilistic attack detection introduces additional randomness into the system dynamics. Our key observation to overcome these issues is that the system dynamics is Markovian and the attacker's injections at any time can be computed based on the knowledge about it, which captures the impact of all the past detection results. To summarize, the main contributions of our work are as follows:

$\bullet$ We solve the aforementioned attacker's problem using a Markov decision process (MDP) framework. In our formulation, the sequential operations of probabilistic attack detection and mitigation are mapped to the MDP's state transition probabilities. The MDP is solved by state space discretization and using the \emph{value iteration} algorithm \cite{Puterman:1994}.

$\bullet$ However, the value iteration algorithm is computationally expensive, especially for high-dimensional systems. To address this issue, we apply a Q-learning with linear and non-linear (neural networks based) function approximation method to solve the attacker's problem in these systems. 

$\bullet$ Based on the above framework, we consider the problem of designing the detection threshold 
from the defender's perspective. To this end, we derive analytical expressions to quantify the cost of FPs and MDs. Based on these cost functions, the attack detection threshold can be tuned to balance the effects of FPs and MDs depending on the accuracy of the mitigation signal.

$\bullet$ We apply our framework to the power grid voltage control that regulates the pilot bus voltages by adjusting the generators' reactive power outputs. 
The attacker injects false voltage measurements, aiming at deviating the pilot bus voltages.
Extensive simulations using PowerWorld show that the optimal attack sequence 
computed using our approach causes the maximum deviations of the pilot bus voltages.

The rest of the paper is organized as follows. Section~\ref{sec:Related} reviews related work. 
Section~\ref{sec:Sys_Model} describes the system model. Section~\ref{sec:Threat_Model} formulates the research problem. 
Section~\ref{sec:Soln_Methods} describes the MDP-based solution. 
Section~\ref{sec:MD_FP} analyzes the impact of FPs and MDs on the system performance. 
Section~\ref{sec:Sim_Res} presents the simulation results. Section~\ref{sec:Conclusion} concludes.

\section{Related Work}
\label{sec:Related}
Most of the existing studies treat attack detection and mitigation problems separately. 
In the category of attack detection, the performance degradation caused by stealthy attacks in a noiseless LTI system has been analyzed \cite{Pasqualetti2013}. However, non-determinism and measurement noises experienced by real-world systems provide an opportunity for the attacker to masquerade his attack as natural noises, thereby rendering attack detection probabilistic. For a general LTI system, research \cite{Mo2015} and \cite{Kwon2013} studied 
the impact of stealthy FDI attacks, and derived optimal attack sequences that can cause the worst system performance degradation. The impact of such stealthy FDI attacks on system efficiency and safety in the context of power grids have also been investigated \cite{Sinopoli_MarketOp2011, RenLoadRedis2011, RuiAGC}. 
In particular, the economic impact of FDI attacks were studied in \cite{Sinopoli_MarketOp2011} and \cite{RenLoadRedis2011}. Reference \cite{RuiAGC} showed that the attacker can drive the power system frequency to unsafe levels by injecting a sequence of carefully-crafted FDI attacks.
However, all these studies \cite{Kwon2013,Mo2015,BaiGupta2014,Sinopoli_MarketOp2011,RenLoadRedis2011,RuiAGC} ignore the attack mitigation triggered by probabilistic detection of attacks and its impact on the future system states and attack detection.

In the category of attack mitigation, preventive and reactive mitigation strategies have been proposed. Preventive mitigation identifies vulnerabilities in the system design and removes them to prevent exploitation by attackers. For instance, in a power grid, a set of sensors and their data links can be strategically selected and protected such that a bad data detection mechanism cannot be bypassed by FDI attacks against other unprotected sensors and links \cite{Bobba2010,Dan2010,KimPoorProtection2011}. However, preventive mitigation provides static solutions only, which do not address the adaptability of strategic attackers against critical infrastructures. Thus, in addition to preventative mitigation, it is important to develop reactive attack mitigation, i.e., countermeasures that are initiated after detecting an attack. Reactive attack mitigation is mainly studied under game-theoretic settings, both in the context of general LTI systems \cite{Barreto2013,ZhuBasar2015} and power grids \cite{EsmaTSG2013,SanjabTSG2016,Ma2013}.
In particular,  \cite{EsmaTSG2013} and \cite{SanjabTSG2016} studied the mitigation of FDI attacks against power grids as one-shot games, where as \cite{Ma2013} formulated the repeated interactions between the attacker and the defender's actions under a Markov game framework. 
However, the aforementioned studies on reactive mitigation assume that the attack has been detected, and ignore the impact of uncertain attack detection on the overall attack mitigation.
In contrast, our framework captures the interdependence between the 
attack detection and mitigation, and their joint impact on the system's dynamics and performance.

\section{Preliminaries} 
\label{sec:Sys_Model}
The analysis in this paper is based on the general discrete-time LTI model. Using this LTI model, we can develop a general framework that applies to many control loops in the power grid (please refer to Section~\ref{subsubsec:lti-power}), rather than being specific to individual control loops and analyzing each of them separately.
\subsection{System Model}
A block diagram of the system model is illustrated in Fig.~\ref{fig:sys_model}.
We consider a general discrete-time LTI system evolving as
\begin{align}
\xv[t+1] &= \Am \xv[t] + \Bm \uv[t] + \wv[t], \label{eqn:process}
\end{align}
where $\xv[t] \in \mathbb{R}^{n}$ is the system state vector, 
$\uv[t] \in \mathbb{R}^{p}$ is the control input, and $\wv[t] \in \mathbb{R}^{n}$ is the process noise at the $t$-th time slot. Matrices $\Am$ and $\Bm$ denote the propagation and control matrices, respectively.
The initial system state $\xv[0]$ and process noise $\wv[t]$ are independent Gaussian random variables. Specifically, $\xv[0] \sim \mathcal{N} (\bf{0},\Xm)$ and $\wv[t] \sim \mathcal{N} (\bf{0},\Qm),$ where $\mathbf{0} = [0, \ldots, 0]^T$ and $\mathbf{X}$ and $\mathbf{Q}$ are the covariance matrices.
The process described in \eqref{eqn:process} is observed through sensors deployed in the system, whose
observation at time $t$, denoted by $\yv[t] \in  \mathbb{R}^m$, is given by
\begin{align}
\yv[t] &= \Cm \xv[t] + \vv[t], \label{eqn:Obs}
\end{align}
where $\Cm \in  \mathbb{R}^{m \times n}$ is the measurement matrix and $\vv[t] \sim \mathcal{N} (\bf{0},\Rm)$ is the measurement noise at time $t$ with a covariance of $\Rm$. We assume that $\vv[t]$ is independent of $\xv[0]$ and $\wv[t]$. Moreover, we assume that the system in \eqref{eqn:process} is controllable and the measurement process in \eqref{eqn:Obs} is observable. 

From a practical point of view, the process noise models the mismatch between the actual system state (following a control action) and that predicted by the linear model in \eqref{eqn:process}. The observation noise is due to the sensor measurement noise. In the context of power grids, these noises can be modeled as Gaussian random variables\footnote{Although our analysis assumes Gaussian noise, in Section~\ref{sec:Sim_Res}, we perform simulations with non-Gaussian noises and show the performance of the algorithms developed under this setting. The non-Gaussian setting may be important to model real-world sensor measurement noises, for instance, noises from phasor measurement units (PMUs)  \cite{Wang2018}. }.

\subsubsection{LTI Systems in Power Grids}
\label{subsubsec:lti-power}
\begin{figure}[!t]
\centering
\includegraphics[width=0.45\textwidth]{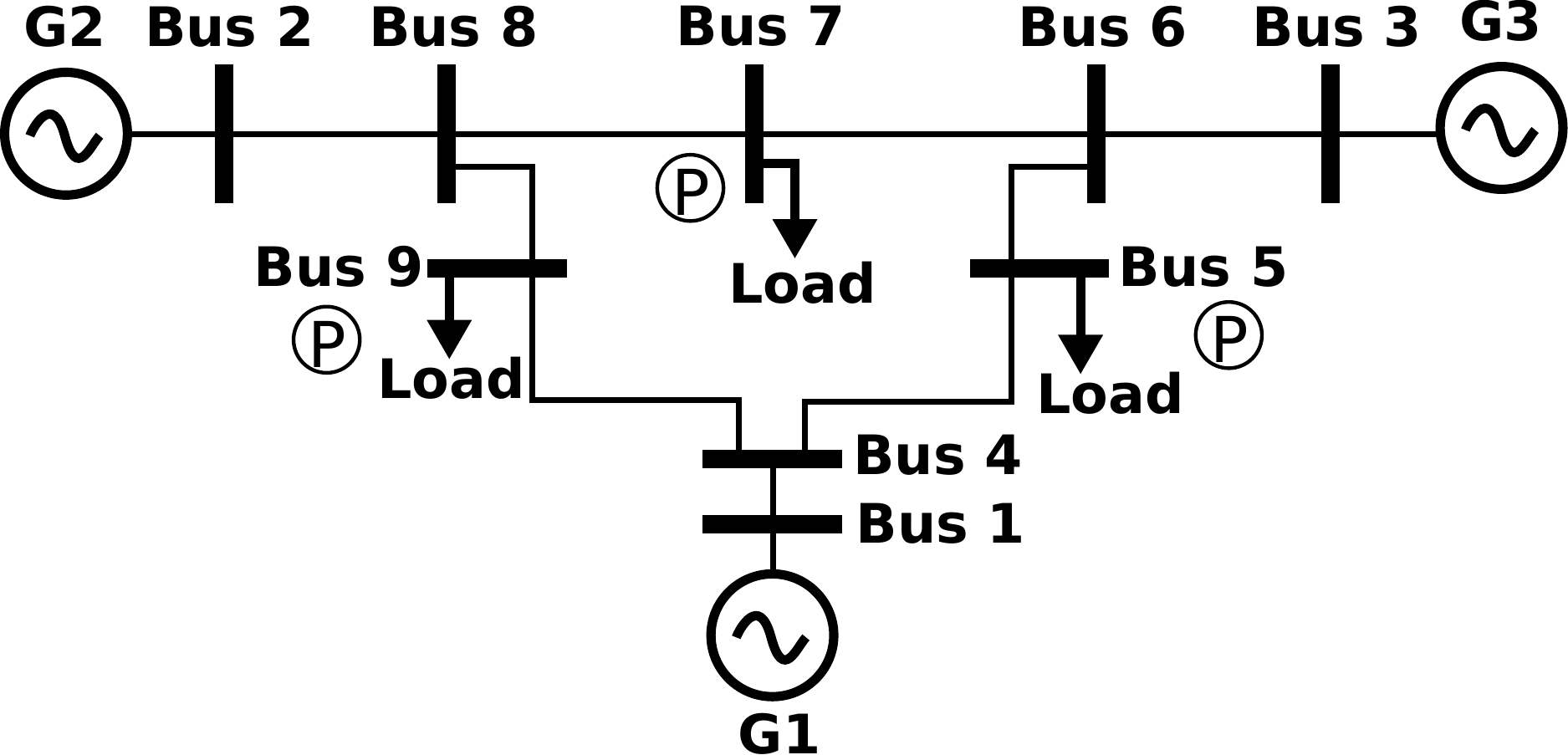}
\caption{IEEE 9-bus power system.}
\label{fig:9_bus}
\end{figure}

We provide examples of a discrete-time LTI system, namely a power system's voltage control and generator swing equations. A power system consists of a set of buses (nodes) to which generators and loads are connected, and transmission lines connecting these buses. Fig.~\ref{fig:9_bus} illustrates the IEEE $9$-bus test system.

\emph{Voltage control}: Power system voltage control refers to maintaining the voltages of selected critical buses (called {\em pilot buses} marked with ``P" in Fig.~\ref{fig:9_bus}) within safe operational limits by adjusting the 
output voltage of the generator buses \cite{Paul1987, Ilic1995}. It can be modeled as an LTI system described in Eqs. \eqref{eqn:process} and \eqref{eqn:Obs}. Specifically, the state vector $\xv[t]$ refers
to the voltages of the pilot buses at time $t,$ which should be maintained at a nominal
voltage denoted by $\xv_0.$ The control signal, which is applied at the generator buses, corresponds to the change in the generator bus voltages, i.e., $\uv[t] = \vv_G[t]-\vv_G[t-1],$ where $\vv_G[t]$ is a vector of the generator bus voltages. We clarify here that we do not consider the detailed model of the generator control loop at the generator bus explicitly. Instead, the control actions in our model only considers the changes in the voltage of the generator bus, while abstracting away finer details of the generator control loop at the generator buses. 

Under this model, the voltage control system can be approximated by an LTI system with $\Am = \Id$ \cite{Ilic1995},\cite{Paul1987}. The control matrix $\Bm$ models the dependency between change of voltage at the generator bus (i.e., the control signal) and the change of voltage at the pilot buses (i.e., the state). This relationship depends on the power flow within the network. Note that in our work, instead of using the power flow equations to model $\Bm$ (e.g., by ac or dc power flow equations), we used a data-driven approach in which we estimate the matrix $\Bm$ from data traces obtained in a PowerWorld simulation. More details on estimating the matrix $\Bm$  will be presented in Section~\ref{sec:Sim_Res}. 
Since the estimation cannot be perfect, the LTI model may be inaccurate, though the inaccuracies are small and can be captured as  process noise. 
As the system state
can be directly measured by voltage sensors (e.g., phasor measurement units, PMUs) deployed at the pilot buses, the measurement matrix is an identity matrix, i.e., $\Cm = \Id.$
The system is bounded-input bounded-output stable if the control algorithm 
satisfies $\Bm \uv[t] = \alpha (\xv_0 - \xv[t])$ for $\alpha \in (0,1)$, and this control
is adopted in practical systems \cite{Paul1987}. However, as the sensor measurements are noisy, the controller cannot have perfect knowledge of the system state $\xv[t].$ Rather, the state is estimated
using the KF-based technique described in \eqref{eqn:KF_est_nomit}.
Based on the estimated state $\hat{\xv}[t],$ the control can be computed as $\uv[t] = \alpha \Bm^{-1} (\xv_0 - \hat{\xv}[t])$.

\emph{Generator swing equations:} The swing equations establish a mathematical relationship between the angles of the mechanical motor and the generated alternating current electricity \cite{Kundur1994}. The swing equations can be linearized and modeled as an LTI system described by Eqs. \eqref{eqn:process} and \eqref{eqn:Obs} under the assumption of direct current (DC) power flow. For a power network consisting of $n$ generators, the state vector consists of $2n$ entries. The first $n$ entries are the generator's rotor phase angles and
the last $n$ entries are the generator's rotor frequency. The control inputs correspond to 
changes in mechanical input power to the
generators, and is responsible for maintaining the generator's rotor angle and frequency within a safe operational range.
The entries of the matrix $\Am$ depend on the power system's topology (including the transmission lines' susceptances) 
as well as the generators' mechanical parameters (such as inertia and damping constants). 
The structure of the matrix $\Bm$ depends on the type of feedback control used to restrict the rotor 
angle frequency to within the safety range \cite{Kundur1994}. 
The measurement vector $\yv[t]$ under the DC power flow model includes nodal real power
injections at all the buses, all the branch power flows, and the rotor angles. The observation matrix $\Cm$ can be constructed based on the power system topology \cite{wood1996power, MonticelliStEst2000}. These linearized control models have been widely used in power grid research literature \cite{FabioPow2011, Zhao2014, Amini2018}.

\subsubsection{Power Grids Dynamic State Estimation Using Kalman Filter}
We consider Kalman filter (KF), which is widely used for state estimation of a dynamic system, and adopted in power grids' transmission \cite{Coutto2009} and distribution grids \cite{Rosenberg2018} as well. The KF works as follows:
\begin{align}
\hat{\xv} [t+1] \!=\! \Am \hat{\xv}[t]\! +\! \Bm \uv[t] \! +\! \Km (\yv [t+1]\! -\! \Cm (\Am \hat{\xv}[t]\! +\! \Bm \uv[t])), \label{eqn:KF_est_nomit}
\end{align}
where $\hat{\xv}[t]$ and $\hat{\xv}[t+1]$ are the estimates of the system states at the $t-$th and $t+1-$th time slots, respectively, $\Km$ denotes the steady-state Kalman gain given by $\Km =  \Pm_\infty \Cm^T(\Cm  \Pm_\infty\Cm^T + \Rm )^{-1}$, and 
the matrix $\Pm_\infty$ is the solution to the algebraic Riccati equation $\Pm_\infty = \Am\Pm_\infty\Am^T + \Qm - \Am\Pm_\infty\Cm^T(\Cm\Pm_\infty\Cm + \Rm)^{-1}\Cm\Pm_\infty\Am^T.$
We denote the KF estimation error at time $t$ by $\ev[t] = \xv[t] - \hat{\xv}[t]$.

Note that in this work, we focus on the estimation of  a sequence of steady states of the power system. Compared to single snapshot-based weighted least square (WLS) method (i.e., the traditional state estimation algorithm \cite{wood1996power}), dynamic state estimators like Kalman filter exploit the valuable historical information of the system states to reduce the computational complexity involved in computing the state estimate (i.e., observe that the evolution of the state estimate is linear \eqref{eqn:KF_est_nomit}).

\begin{figure*}[!t]
\centering
\includegraphics[width=0.62\textwidth]{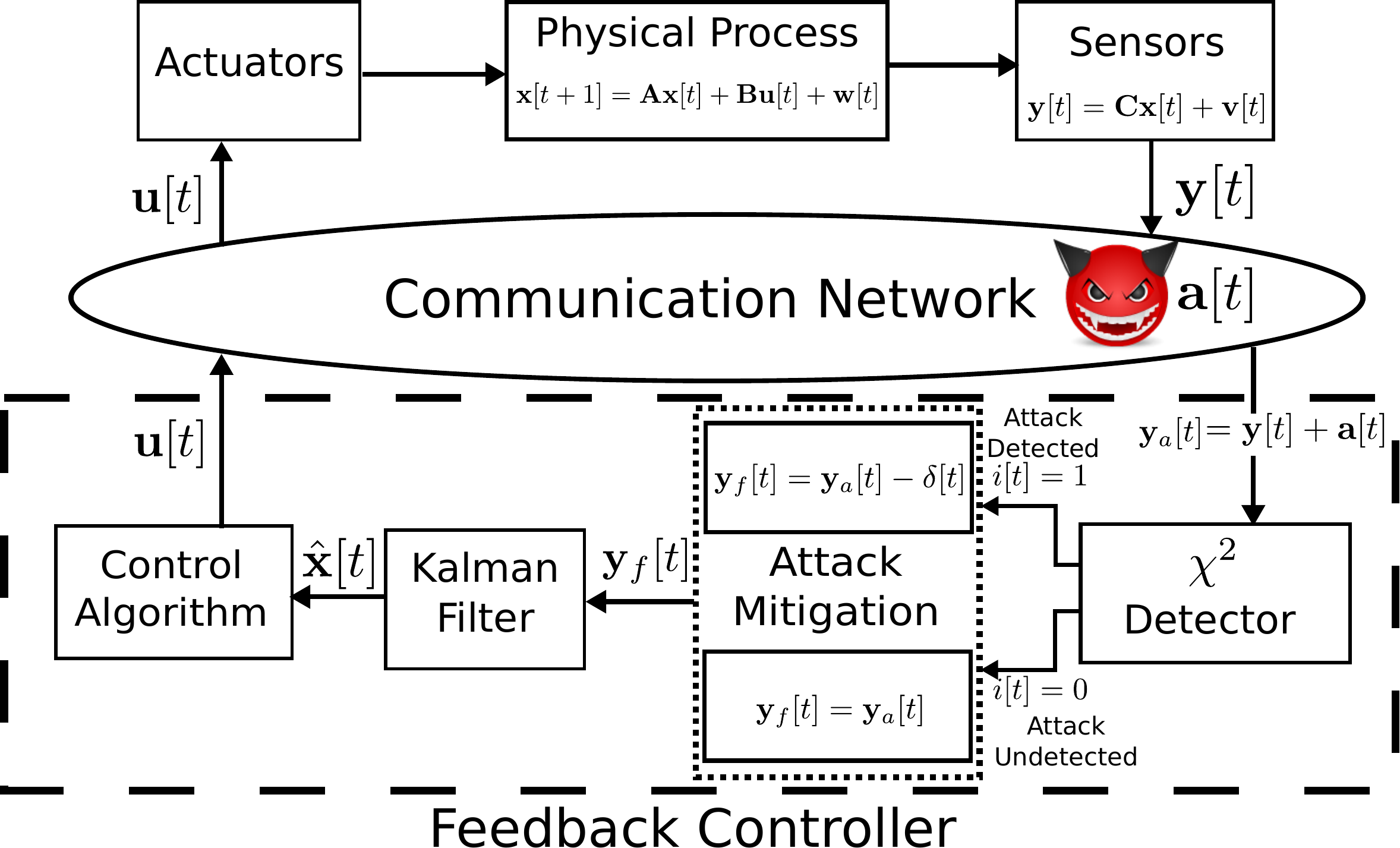}
\caption{Block diagram of the system model.}
\label{fig:sys_model}
\end{figure*}

\subsection{Attack Model, Detection, and Mitigation}
\subsubsection{Attack Model} 
Modern-day critical infrastructure systems extensively use ICT for their operation.
For instance, in a power grid, the remote terminal units (RTUs) and many other field
devices are connected by the Internet protocol (IP) \cite{Shodan}. 
Additionally, in a power grid, the sensors (such as the voltage and current measurement units) are spread over a large geographical area, making their measurements vulnerable to physical attacks. Such vulnerabilities can be exploited to launch attacks and
disrupt the normal power grid operations.

In this paper, we follow Kerckhoffs's principle and consider an 
attacker who has accurate knowledge of the targeted CPCS and read access to the system state.
Such knowledge can be obtained in practice by malicious insiders, long-term
data exfiltration \cite{dragonfly2014}, or social engineering against employees,
contractors, or vendors of a critical infrastructure operator \cite{karnouskos2011}.
Specifically, we assume that the attacker knows the matrices $\Am, \Bm$ and $\Cm,$ as well as the operational details of the KF and the system's method of anomaly detection (including the detection threshold). In addition, the attacker also has read and write accesses to the system measurements, e.g., using PLC rootkits \cite{GarciaPLC2017}.

We consider FDI attacks on the system sensors. The compromised
observations, denoted by $\yv_a[t]$, are given by $\yv_a[t] = \yv[t] + \av[t]$, where $\av[t] \in  \mathbb{R}^{m}$ is the attacker's injection.
To model the attacker's energy constraint, we assume that the norm of the injection, $\|\mathbf{a}[t]\|$, is upper-bounded by a constant $a_{\max}$, i.e., $||\av[t]|| \leq a_{\max}.$
Denote by $\mathcal{A}$ the set of all feasible attack vectors that satisfy the above energy constraint. 

\subsubsection{Attack Detection and Mitigation}
We assume that the controller uses the $\chi^2$ detector \cite{MEHRAChiSquare1971} to detect the attack, which
has been widely adopted in security analysis of LTI systems \cite{Kwon2013}, \cite{Mo2015}. The $\chi^2$ detector computes a
quantity
$g[t] = \rv[t]^T \Pm^{-1}_r \rv[t],$
where $\rv[t]$ is the residual given by
$\rv[t+1] = \yv_a[t+1] - \Cm (\Am \hat{\xv}[t] + \Bm \uv[t])$
and $\Pm_r = \Cm \Pm_{\infty} \Cm+\Rm$ is a constant matrix that denotes the covariance of the residual in the steady state. The detection result of the detector is denoted by $i[t] \in \{0, 1\}$. The detector declares an attack if $g[t]$ is greater than a predefined threshold $\eta.$ 
Specifically, if $0 \leq g[t] \leq \eta$, $i[t]=0$; otherwise, $i[t]=1$.

Based on the detection result, the controller applies a reactive mitigation action. 
If the $\chi^2$ detector's alarm is triggered, the controller forwards a modified version of the observation $\yv_a[t] - \deltav[t]$  
to the KF, where $\deltav[t] \in  \mathbb{R}^m$ is an attack mitigation signal;
otherwise, the controller directly forwards $\yv_a[t]$ to the KF (ref. Fig.~\ref{fig:sys_model}). 
Thus, the controller's operation is
\begin{align}
\yv_f[t] = \yv_a[t] - i[t] \deltav[t]. \label{eqn:mitigation}
\end{align}
With the controller's mitigation action, the KF estimate is
\begin{align}
\hat{\xv} [t\!\!+\!\!1] \!\!=\!\! \Am \hat{\xv}[t]\!\! +\!\! \Bm \uv[t] \!\! +\!\! \Km (\yv_f [t\!\!+\!\!1] \!\! - \!\! \Cm (\Am \hat{\xv}[t] \!\! + \!\! \Bm \uv[t])). \label{eqn:KF_est}
\end{align}

The mitigation signal $\deltav[t]$ can be generated using existing mitigation approaches \cite{Cardenas2011,Sridhar2014, FawziTAC2014,Mishra2017}. Specifically, once the attack is detected, a conservative mitigation approach is to ignore the sensor measurements completely, and drive the system based only on the model. For instance, the estimated state $
\widehat{\xv}[t+1] = \Am \widehat{\xv}[t] + \Bm \uv[t]$ can be which is subsequently used to compute the control decisions.
Alternately, upon detecting an attack, the controller can ignore the sensor measurements, and instead make an \emph{educated guess} to obtain the true sensor measurements \cite{Sridhar2014}. For instance, in power grids, accurate real-time load forecasts can be computed using techniques such as regression models, neural networks and statistical learning algorithms. The forecasted measurements can be used to drive the Kalman filter estimates.
The operator can also run a secure state estimation algorithm (e.g., \cite{FawziTAC2014,Mishra2017}) to obtain a reliable estimation of the system state in the presence of sensor attacks.

The main focus of this paper is not the design of the mitigation strategy, but to understand 
the impact of the detection-mitigation loop on the optimal attack strategy.
Thus, in this paper, we do not focus on a specific mitigation approach. Instead, we design a generic framework 
that admits any mitigation signal.
In Section~\ref{sec:Sim_Res}, our simulations are based on a perfect mitigation strategy in which the
controller can precisely remove the attack signal, as well as a practical mitigation strategy in which the mitigation signal is a noisy version of the attack signal.

Combining \eqref{eqn:process}, \eqref{eqn:Obs}, \eqref{eqn:mitigation} and \eqref{eqn:KF_est}, 
we obtain the dynamics of the KF estimation error with attack mitigation as
\begin{multline}
\ev  [t+1]  =  \Am_K \ev[t] + \Wm_K\wv [t]   -  \Km \av_m[t+1] - \Km \vv[t+1], \label{eqn:error_evol}
\end{multline}
where $\av_m[t+1] = \av[t+1]-i[t+1] \deltav[t+1], \Am_K = \Am- \Km \Cm \Am$ and $\Wm_K = (\Id- \Km \Cm).$ Further, we have $\mathbb{E}[\ev[0]] = \bf{0}$ and $\mathbb{E} [\ev[0] \ev[0]^T] = \Pm_e = (\Id- \Km \Cm) \Pm_\infty.$

\section{Problem Formulation}
\label{sec:Threat_Model}
Under the Kerckhoffs's assumption about the attacker's
knowledge, we analyze attack strategies that can mislead the controller into making erroneous control
decisions. This is accomplished indirectly by 
increasing the estimation errors. 
For a given attack detection threshold $\eta$ and mitigation strategy $\{ \deltav[t] \}^T_{t = 1}$ over a horizon of $T$ time slots, the optimal attack sequence that maximizes the cumulative sum of KF's expected square of norm of the estimation error over the horizon is given by the following optimization problem:
\beqa
&\dsp  \max_{ \av[1],\dots,\av[T]} &  \sum^T_{t = 1}\mathbb{E} [\| \ev[t] \|^2 ] \label{eqn:attacker_problem} \\ 
& s.t. &  \text{KF error dynamics \eqref{eqn:error_evol}}, \|\mathbf{a}[t] \| \leq a_{\max}, \forall t \nonumber.
\eeqa 
Maximizing the KF estimation error implies that the controller no longer has
an accurate estimate of the system state. 
In systems that use KF for state estimation (such as power grid's voltage control etc.), control input computed based on inaccurate/wrong system state estimates can adversely affect their performance and even result in catastrophic safety incidents. 
Moreover, the cumulative sum in the objective function implies that the attack has a sustained adverse impact on the system over the entire attack time horizon. 
We note that similar cumulative metrics have also been widely adopted in control system design to assess the performance of controllers \cite{abdelzaher2008introduction}.

\begin{figure}[!t]
\centering
\includegraphics[width=0.48\textwidth]{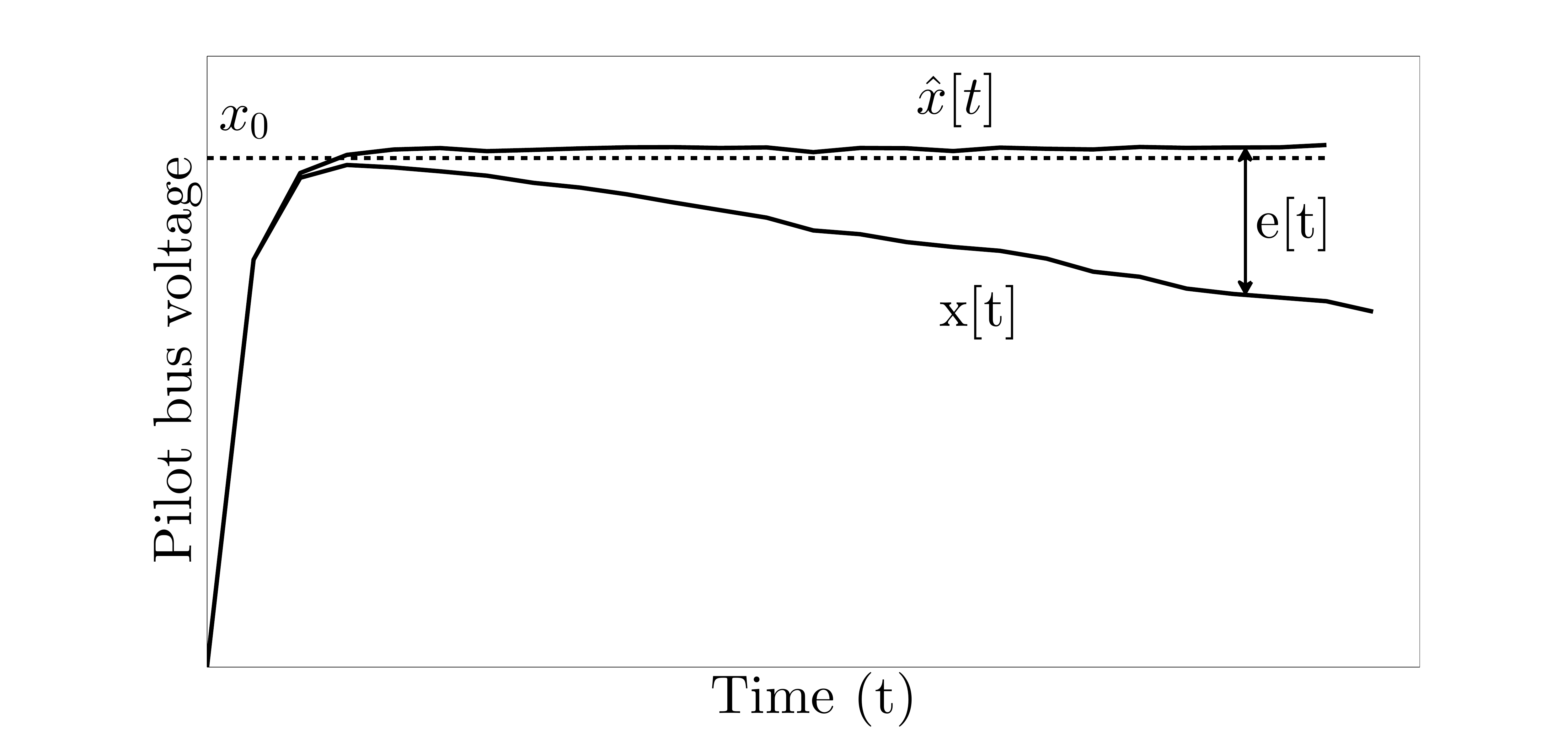}
\caption{Attack impact for the voltage control problem.}
\label{fig:obj_fn}
\end{figure}

We now illustrate the relevance of \eqref{eqn:attacker_problem} to power grid's voltage control. 
Fig.~\ref{fig:obj_fn} shows the impact of an attack that is able to bypass the  $\chi^2$ detector (and consequently the controller's mitigation steps) on the pilot bus voltage. 
In this figure, the dotted line indicates the voltage setpoint, and the solid lines show the evolution of the system state $\xv[t]$ and estimate $\hat{\xv}[t]$. The gap between the two curves measures the KF estimation error $\ev[t].$ As evident from the figure, if the attacker manages to increase the KF's estimation error using a carefully constructed attack sequence, he can cause a significant deviation of the system state from the desired setpoint. Interestingly, the estimate $\hat{\xv}[t]$ is close to the setpoint ${\xv}_0$ that misleads the controller into believing that the desired setpoint has already been achieved, while the actual pilot bus voltage continues to deviate.

Intuitively, to cause a significant impact, the attack magnitude must be large. But at the same time, it is important that the attack bypasses the controller's detection -- otherwise the attack will be mitigated. Thus, the solution of the optimization problem \eqref{eqn:attacker_problem} must strike a balance between the attack magnitude and stealthiness. 
In the following section, we solve the optimization problem \eqref{eqn:attacker_problem} 
using an MDP-based approach.

\section{MDP Solution}
\label{sec:Soln_Methods}
In this section, we cast the optimization problem \eqref{eqn:attacker_problem} to an MDP problem \cite{Puterman:1994} and solve it using value iteration.

{\bf Markov Decision Process Model:} A state in the MDP corresponds to the
KF filter estimation error $\ev[t]$ and the actions correspond 
to the attacker's injection $\av[t].$ 
Our approach is to map the KF error dynamics \eqref{eqn:error_evol} to the
state transition probabilities of the MDP, and the objective function of \eqref{eqn:attacker_problem} to the MDP's long-term expected reward. 
The mathematical details of the MDP is presented next.

Formally, the MDP is defined by a tuple $(\mathcal{E},\mathcal{A},\mathcal{T},R)$, where
$\mathcal{E} \subseteq \mathbb{R}^n$ is the state space of the problem corresponding 
to the set of all possible $\ev[t]$.  $\mathcal{A}$ is the action space of the attacker; $\mathcal{T}(\ev,\av,\ev^{\prime})$ is the probability of transiting from state
$\ev$ to $\ev^{\prime}$ (where $\ev,\ev^{\prime} \in \mathcal{E}$) under an action $\av \in \mathcal{A}$ of the attacker; formally,
$\mathcal{T}( \ev,\av,\ev^{\prime})  \doteq \mathbb{P}  (\ev[t+1] = \ev^{\prime} \big{|} \ev[t] = \ev,\av[t+1] = \av) .$
$R(\ev^{\prime},\av,\ev)$ is the immediate expected reward for the attacker when it takes an action 
$\av \in \mathcal{A}$ in state $\ev \in \mathcal{E}.$

\emph{MDP state transition probabilities:} We adopt the following approach. First, we compute the quantity $\mathbb{P} ( \ev_{\text{lb}} \leq \ev[t+1]  \leq \ev_{\text{ub}} \big{|} \ev[t] = \ev,\av[t+1] = \av),$ for 
any $ \ev_{\text{lb}} , \ev_{\text{ub}} \in \mathbb{R}^n$ and $\ev_{\text{lb}} \leq \ev_{\text{ub}}$ from \eqref{eqn:error_evol}. Then, we use the fact that for a random variable $X,$ 
$\mathbb{P}  (X = x) \approx \frac{F ( -\infty ,x+\epsilon) - F ( -\infty ,x-\epsilon) }{2 \epsilon},$
where $F ( x_1 ,x_2) = \mathbb{P} (  x_1 \leq X  \leq x_2)$ and $\epsilon > 0$ is a small positive quantity. 
The result is stated in the following lemma:
\begin{lemma}
\label{lem:trans_prob}
For a given $\ev[t] = \ev$ and $\av[t+1] = \av$ the quantity $\mathbb{P} ( \ev_{\text{lb}} \leq \ev[t+1]  \leq \ev_{\text{ub}} \big{|} \ev[t] = \ev,\av[t+1] = \av)$  can be computed as the sum of the following terms:

\begin{small}
\begin{align}
& \mathbb{P} \LB \begin{bmatrix}
 0  \\
 \ev_{\text{lb}} - \yv_2
\end{bmatrix}  \leq \Xm   \leq \begin{bmatrix}
 \eta  \\
 \ev_{\text{ub}} - \yv_2
\end{bmatrix} \RB  \nonumber \\ & + \mathbb{P} \LB \begin{bmatrix}
 \eta  \\
 \ev_{\text{lb}} - \yv_2 - \Km \deltav
\end{bmatrix} 
 \leq \Xm   \leq \begin{bmatrix}
 \infty  \\
 \ev_{\text{ub}} - \yv_2 - \Km \deltav
\end{bmatrix} \RB.
\label{eqn:TP_vector}
\end{align}
\end{small}
In \eqref{eqn:TP_vector}, $\Xm \in  \mathbb{R}^{n+1}$ is a concatenated variable given by 
$\Xm  = \LSB \Ym \ \ (\Wm_K \wv[t]- \Km \vv[t+1] )^T \RSB^T,$ $\yv_2 = \Am_K\ev - \Km\av,$
and $\deltav$ is the mitigation signal.
\end{lemma} 
The proof of Lemma~\ref{lem:trans_prob} is omitted here and presented in Appendix~A of \cite{LaksheEnergy2017}.

\emph{MDP reward:} We now map the objective function of \eqref{eqn:attacker_problem} to the 
MDP reward function. Accordingly, the immediate expected reward of the MDP is given by
$R(\ev ,\av, \ev^{\prime}) = \int_{\ev^{\prime} \in \mathcal{E}}\mathcal{T}( \ev,\av,\ev^{\prime}) || \ev^{\prime} ||^2$.
\label{eqn:imm_rew}

\emph{MDP policy and state value function: }The solution to the MDP corresponds to a policy $\pi,$ which is a mapping from a state to an action. 
The state value function of the MDP for a given policy $\pi$ is defined as
$V^{\pi} (\ev) = \mathbb{E}_{\pi} \LSB \sum^T_{t = 1} ||\ev[t]||^2 \big{|} \ev[0] = \ev \RSB$.

\emph{Optimal policy: }The optimal policy
$\pi^*$ maximizes the total expected reward, $
\pi^* = \argmax_{\pi} V^{\pi} (\ev) , \forall \ev \in \mathcal{E},$ and
the optimal value function is defined as $V^*(\ev) = V^{\pi^*} (\ev).$

{\bf Solving the MDP:} MDPs can be solved by value/policy iteration methods for discrete systems \cite{Puterman:1994}. However, we study discrete-time CPCS with continuous system states. For instance, the voltages in the voltage control system are continuous variables. 
Hence, the MDP described does not admit the value iteration method directly. To address this issue, we define a \emph{discretized MDP} by discretizing the state space of the original continuous MDP. The discretized MDP admits a deterministic optimal policy (i.e., a pure strategy) \cite{Puterman:1994}, (Theorem~4.4.2), which be can computed using the value iteration method (the convergence of value iteration follows from the fact that the updates of the value iteration algorithm form a contraction mapping \cite{Puterman:1994}, Theorem~6.3.1). The optimal policy of the discretized MDP can then be 
used as a near-optimal solution to the continuous MDP \cite{ChowTsitsiklis1991}. (Please also refer to our subsequent arguments in bullet point 2.) The details of the discretization procedure are omitted here and can be found in Appendix~B of reference \cite{LaksheEnergy2017}. 
We make some remarks on the MDP formulation:

\begin{itemize}

\item[1.] Note that to execute the MDP policy, the attacker requires the knowledge of the Kalman filter estimation error $\ev[t],$ which in turn depends on the system state $\xv[t].$ The attacker can use a separate Kalman filter to track $\xv[t]$, since he can access the original sensor measurements.
To simplify our analysis, we assume that the attacker can directly observe $\xv(t)$ and $\ev(t)$. Under this assumption, the attacker is advantageous. Thus, our analysis is conservative from the perspective of the system defender, which is often acceptable for safety-critical systems. We verify the validity of this assumption using simulations for voltage control in the IEEE-9 bus system and present the result in Section~\ref{sec:Sim_Res}.

\item[2.] The optimal cost of the discretized MDP 
is guaranteed to lie within a bounded distance from the optimal cost of the original MDP 
\cite{ChowTsitsiklis1991}. 
As the discretization is finer, the discretized MDP approaches to the original
MDP more closely.

\item[3.] Let $A$ and $D$ denote the size of the action and the state space of the discretized MDP.
The computational complexity of implementing the value iteration algorithm is $\mathcal{O} (A D)$ \cite{Puterman:1994}. We note that the number of discretization levels of the state space $D$ is an important factor. Moreover, the algorithm requires the value function to be stored for each discretized state, making it challenging to implement in high-dimensional systems. 
To address this issue, we present a Q-learning algorithm with function approximation in Section~\ref{sec:Q_learn}, which is more suited
for application in these systems. We also compare the computational overhead of the two algorithms. 

\end{itemize}

\subsection{Attack Magnitude and Stealthiness}
\label{sec:Mag_Stl}
\begin{figure}[!t]
  \centering
  \includegraphics[width=0.23\textwidth]{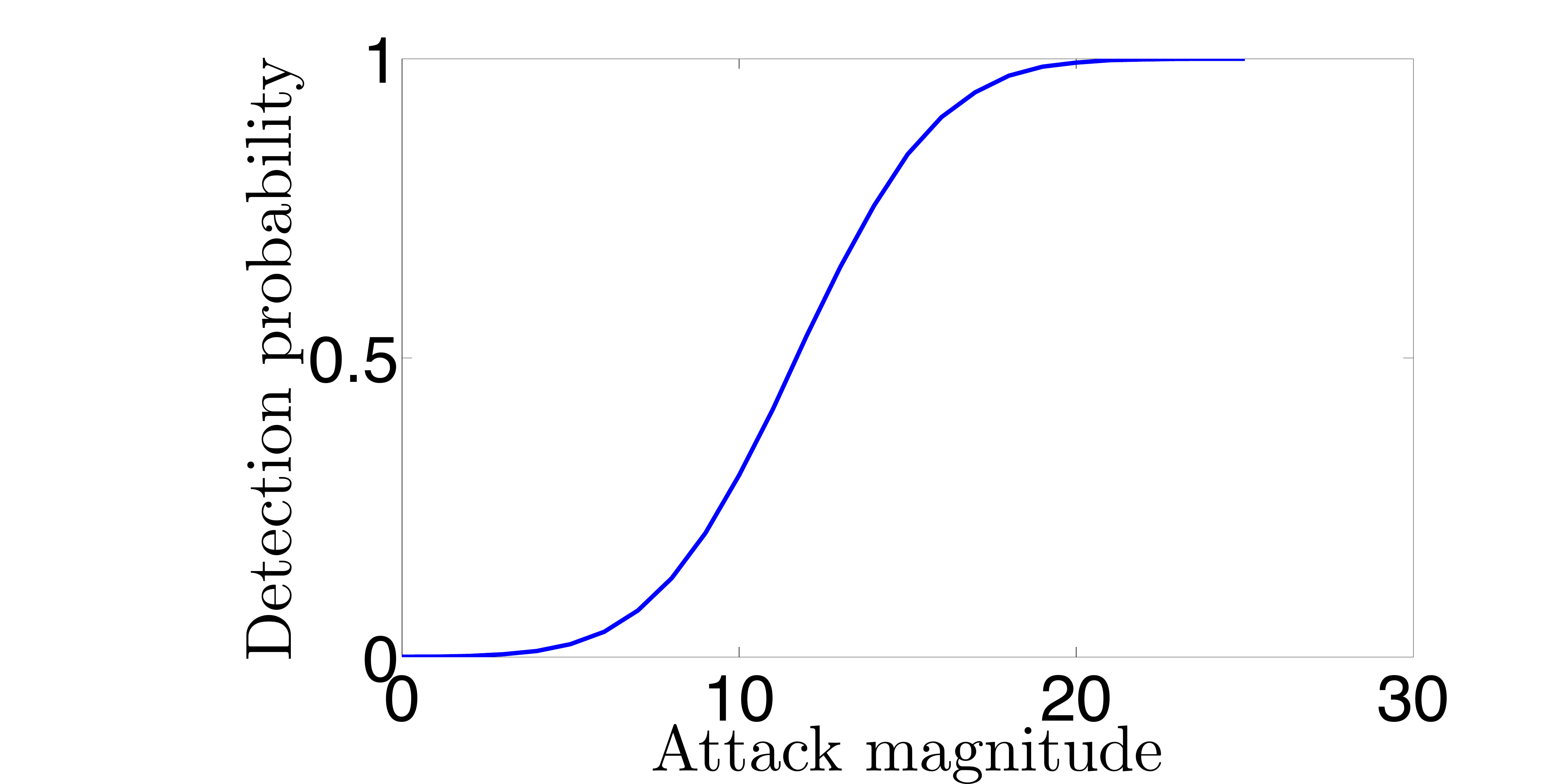}
  \hfill
  \includegraphics[width=0.23\textwidth]{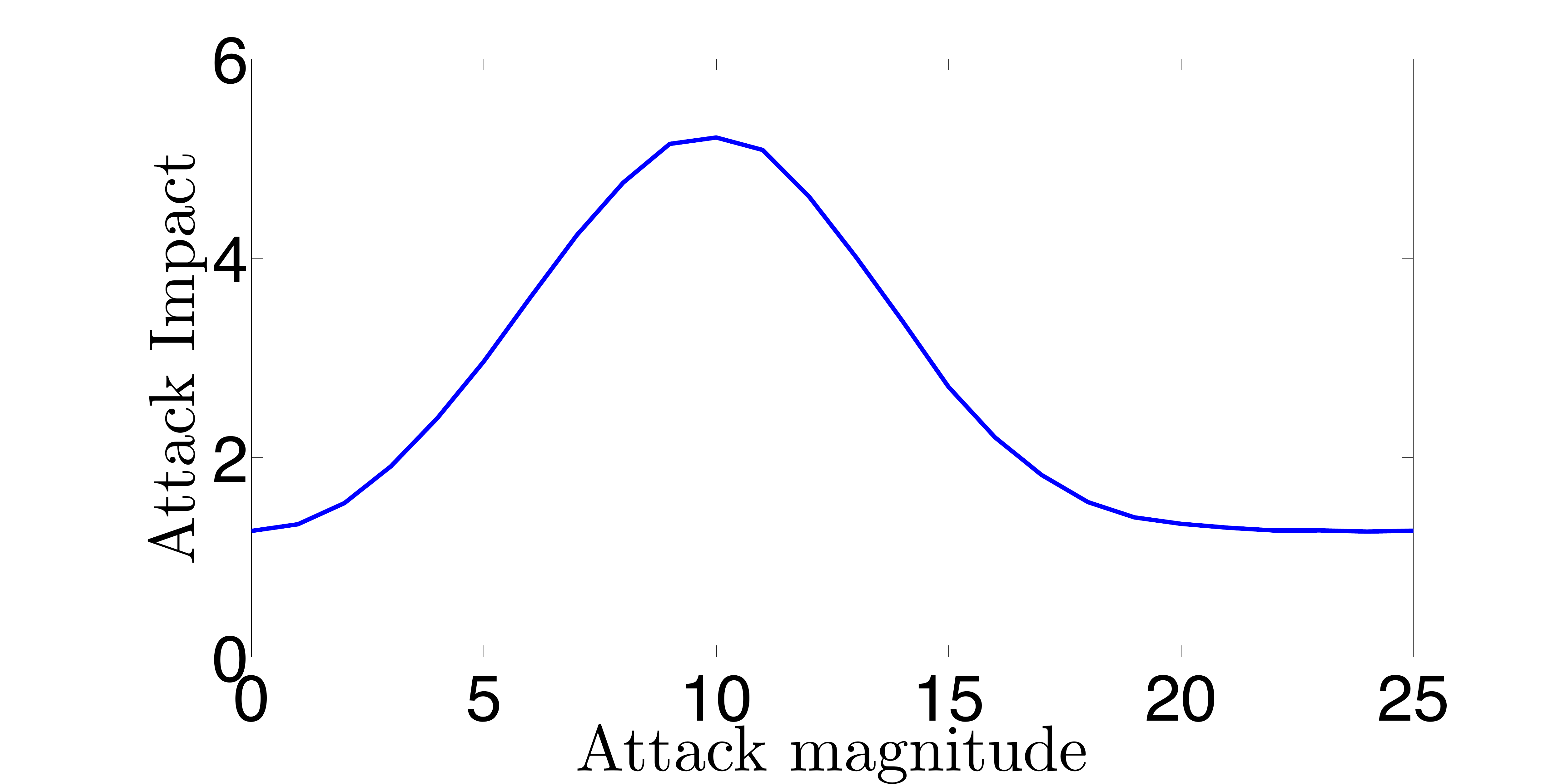}
\vspace{-1em}
\caption{Attack detection probability and the expected attack impact (immediate expected reward of MDP) for different attack magnitudes.} 
\label{fig:Attack_Mag_Stl}
\vspace{-0.5 cm}
\end{figure}

We now illustrate the structure of the MDP solution using a numerical example.
In Fig.~\ref{fig:Attack_Mag_Stl}, we plot the attack detection probability and the attack impact computed in terms of the MDP's immediate expected reward for different values of attack magnitude $\av$. The system parameters are $n = m = 1,$ $\Am$ = $1$, $\Cm$ = $1$, $\Qm$ = $1$, $\Rm$ = $10,$  $\eta = 10$ and $\deltav = \av$. It can be observed that while the probability of detection is low for an attack of small magnitude, it also has little impact. On the other hand, the probability of detection is high for an attack of large magnitude, and consequently the expected attack impact is also low. The optimal attack lies in between these two quantities. In this example, the optimal attack that maximizes the expected immediate reward has a magnitude of $10,$ and 
a detection probability of $0.3.$ Thus, the MDP solution strikes a balance
between attack magnitude and stealthiness, resulting in maximum impact\footnote{Strictly speaking, MDP solution maximizes the long term expected reward. For the ease of illustration, in this example we only considered the immediate expected reward.}.

\section{Costs of False Positives And Missed Detections}
\label{sec:MD_FP}
In this section, we use the framework established in Section~\ref{sec:Threat_Model} and Section~\ref{sec:Soln_Methods} to quantify 
the costs of false positives and missed detections. We use the cumulative state estimation error (objective function of \eqref{eqn:attacker_problem}) as the cost metric.
To quantify these costs, we consider an LTI system with an \emph{oracle} attack detector 
as the reference system. The oracle detector achieves perfect detection capability, that is, it generates no FPs and MDs. The cost of FPs is then the additional cost incurred, compared with the reference system, by wrongly triggered mitigations in the LTI system of Fig.~\ref{fig:sys_model} (defined by \eqref{eqn:process} and \eqref{eqn:Obs} with the $\chi^2$ detector and mitigation modules), compared with the reference system. The cost of MDs is the additional cost incurred by unmitigated attacks in the LTI system of Fig.~\ref{fig:sys_model}.

{\bf Cost of FPs:} To quantify the cost of FPs, we compute the state estimation error (objective function of \eqref{eqn:attacker_problem}) in the following two systems: (i) the LTI system of Fig.~\ref{fig:sys_model} under no attacks, i.e., $\av[t] = 0, \ \forall t;$ (ii) the reference LTI system with $\av[t] = 0, \ \forall t$.
Under setting (i), all the alarms of the $\chi^2$ detector correspond to FPs, which wrongly trigger a mitigation action. Since the mitigation signal is imperfect, it leads to an increase in the estimation error. The difference between the state estimation errors of the two systems quantifies the performance degradation due to FPs.

\vspace*{0.1 cm}
{\bf Cost of MDs:} 
To quantify the cost of MDs, we compute the state estimation errors in the following two systems: 
(i) the LTI system of Fig.~\ref{fig:sys_model} with optimal attacks (ii) the reference LTI system with optimal attacks. The difference between the state estimation errors of the two systems quantifies the performance degradation due to MD, i.e., the cost of MD. In particular, the optimal attacks  derived in Section~\ref{sec:Soln_Methods} characterizes the worst-case performance degradation due to MDs. 

It is hard to characterize the state estimation error (objective function of \eqref{eqn:attacker_problem}) analytically for a generic LTI system. Nevertheless, for $n = m = 1,$ 
we  present a recursive method to compute the state estimation error under any attack sequence $\av[t] , \forall t$. 
We let $\mathcal{B}^t = \{ 0, 1 \}^t$ denote the set of all binary combinations of length $t,$  and 
 $f_{\iv_{[1:t]}} (\bv) = \mathbb{P} ( \iv_{[1:t]} = \bv ), \forall \ \bv \in \mathcal{B}^t$ denote the pdf of $\iv_{[1:t]}.$ Additionally, we introduce the following notations to denote the 
conditional random variables  $\ev_c[t] =  \ev[t]  \big{|} \iv_{[1:t]},$
$\rv_c[t]  = \rv[t] \big{|} \iv_{[1:t-1]},\ev^{(0)}_c[t+1] = \ev[t+1] \big{|}  \{ \iv_{[1:t]}, i[t+1] = 0  \}$ and $\ev^{(1)}_c[t+1] = \ev[t+1] \big{|}  \{ \iv_{[1:t]}, i[t+1] = 1  \}.$ The cumulative state estimation error in the LTI system of Fig.~\ref{fig:sys_model} can be computed as 
$\mathbb{E}[ || \ev[t] ||^2 ]  = \sum_{ \bv \in \mathcal{B}^t }f_{\iv_{[1:t]}} (\bv)  \mathbb{E}[|| \ev_c[t] ||^2]$,
where the terms $\mathbb{E}[||\ev_c[t]||^2]$ can be evaluated recursively as

\begin{small}
\begin{equation}
\mathbb{E}[(\ev^{(i)}_c[t\!+\!1])^2] \!=\! \text{Var} ( \ev^{(i)}_d[t\!+\!1]) \!+\! \frac{\int_{a_i}^{b_i} ((\mathbb{E} \LSB  \ev^{(i)}_d[t\!+\!1]  \RSB)^2 f_{\yv}(\yv))}{ \int_{a_i}^{b_i} f_{\yv}(\yv)},
\label{eqn:mean_ec_final}
\end{equation}
\end{small}
where $i=0,1$. In \eqref{eqn:mean_ec_final}, $\mathbb{E}[\ev^{(i)}_d[t+1] ] $ and  $\text{Var} \LB \ev^{(i)}_d[t+1]  \RB $ are the mean and variance of the variable $\ev^{(i)}_d[t+1].$ They can be computed using the result presented in Appendix~A-Part I. The complete derivation of \eqref{eqn:mean_ec_final} can be found in Appendix~A-Part II.

Using the result above, the cost of FPs can be computed by setting $\av[t] = {\bf 0}, \forall t$ (i.e., no attacks) and the cost of MDs under any non-zero attack sequence $\av[t], \forall t$ by injecting the appropriate attacks. The worst-case cost of MDs under the optimal attacks (computed as in Section~\ref{sec:Soln_Methods}) can be directly derived from the value function $V(\xi_i)$ at the convergence of the value iteration algorithm. 

In Section~\ref{sec:quant_MD_FP}, we present simulation results, based on the derivations in this section, to illustrate the cost of FPs and MDs under different attack detection thresholds and mitigation strategies. We also provide guidelines to tune the attack detection threshold based on this quantification.

\section{Algorithm Implementation in High-Dimensional Systems}
\label{sec:Q_learn}
The value iteration algorithm to solve the MDP (as proposed in Section~\ref{sec:Soln_Methods}) has high computation overhead for high-dimensional systems.
Let $\Xi$ denote the discretized version of the original state space
$\mathcal{E},$ where $\Xi = \{ \xi_1,\dots,\xi_D \},$ 
$D$ is the total number descritization levels. 
Then, the value iteration algorithm loops over each discretized state $\xi_i \in \Xi,$ and the value function $V(\xi_i)$ must be stored for each discretized state $\xi_i \in \Xi$ (refer to Algorithm 1 in reference \cite{LaksheEnergy2017} for details of the value iteration algorithm). For an $n-$dimensional system, if each dimension is discritized into $d$ states, the total number of discretized states is $D = d^n.$ Thus, it becomes challenging to run the value iteration algorithm.  
In this section, we propose 
to use the \emph{Q-learning with function approximation} (QLFA) method \cite{Sutton:1998} to extend the MDP solution to high-dimensional systems.

We first introduce the concept of Q-learning. It is a model-free method for computing the optimal MDP policy that maximizes the time-average reward. The basic idea of Q-learning is 
to perform \emph{trajectory-based sampling}, i.e., run trajectories of the 
Kalman filter estimation error $\langle \ev[t],\av[t],R[t],\ev[t+1],\av[t+1],R[t+1],\dots \rangle$, 
and perform Bellman backups only on the states visited. Thus, the Q-learning algorithm does not loop over each state $\xi_i \in \Xi,$ in contrast to the value iteration algorithm. The Q-learning algorithm computes the action-value function for each state and action pair $Q(\xi_i,\av).$  In particular, for $\xi_i[t] = \xi_i, \av[t] = \av, \ev[t+1] = \xi^\prime_i,$ and $R[t+1] = R,$ the Q-values are updated as
\begin{align}
Q(\xi_i,\av) \!\leftarrow \! (1 \!-\! \alpha) Q(\xi_i,\av) \!+\! \alpha \LB R \!+\! \gamma \max_{\av^\prime} Q(\xi_i^\prime,\av^\prime) \RB, \label{eqn:Q-Learn}
\end{align}
where $\alpha > 0$ is the learning rate. The action $\av[t] = \av$ is chosen according to 
the $\epsilon-$greedy method \cite{Sutton:1998}. The Q-learning algorithm can converge for a finite number of state-action pairs under certain conditions on the learning rate, discount factor, and the reward function \cite{Sutton:1998}.

The Q-learning algorithm still has the issue that it stores the Q-values for each state-action pair, which will be challenging in high-dimensional systems. To address this issue, we propose to combine the Q-learning with the \emph{function approximation} method \cite{Roy2013Tutorial}. The key idea of this approach is to approximate the Q-values using parametric functions instead of storing them individually for each state-action pair. The number of parameters to be stored and updated is much less than the number of states. As a result, this method is more suited for application in high-dimensional systems. In particular, we focus on Q-learning with linear function approximation (QLFA), in which the Q-values are approximated as a linear function of a parameter vector $\thetav(\av), \forall \av $ as  
\begin{align}
Q(\xi_i,\av) = \phiv (\xi_i)^T \thetav (\av) , \ \forall \xi_i \in \Xi, \av \in \mathcal{A}, \label{eqn:Qfromfeats}
\end{align}
where $\phiv(\xi_i)$ is the \emph{feature vector} corresponding to every state $\xi_i.$ The feature vector is a lower-dimensional representation of the state, where the
feature function $\phi: \mathcal{E} \rightarrow \mathbb{R}^{K}$ maps each state-action pair to a feature vector, $K$ being the dimension of the feature vector. The feature function $\phi$ can be constructed for each state using various \emph{basis functions}. For example, one of the simplest basis functions is the \emph{fixed sparse representation} (FSR) with binary encoding. Under FSR, we first discretize the $n-$dimensional state space with uniform grid cells of $d$ levels each. Thus, the state space $\Xi$ has $d^n$ states represented by the tuple $(i,j,\dots) \in \Xi.$ The feature vector corresponding to each state $(i,j,\dots) \in \Xi$ is then constructed as
\begin{align}
\phi ((i,j,\dots,)) = [ \underbrace{ 0, .. , \underbrace{1}_{\text{$i^{th}$ position}}, ..,0 }_{\text{$d$ elements}}, \underbrace{ 0, .. , \underbrace{1}_{\text{$j^{th}$ position}}, ..,0 }_{\text{$d$ elements}},..]^T .
\end{align}
Note that $\phi ((i,j,\dots,))$ is a $nd-$dimensional vector (thus, in this case, $K = nd$).

Other choices of basis functions to construct the feature vector include the use of Radial Basis Functions (RBFs) and Incremental Feature Dependency Discovery (iFDD) \cite{Roy2013Tutorial}. The choice of basis function offers a trade-off between accuracy (of estimating the correct value of the MDP's value function) and learning speed.

Under the function approximation method, finding the optimal policy involves updating the 
parameter vector $\thetav(\av)$ (rather than the Q-values). Thus, the total number of elements to be stored is $nd A$ (as opposed to $DA = d^n A$ for the direct Q-learning algorithm). We specify the QLFA algorithm in the following. 

\begin{algorithm}
  \small
\SetAlgoLined
\KwData{MDP, $\alpha$, $\gamma$}
\KwResult{Policy $\pi$}
$\thetav(\av) \leftarrow$ Initialize arbitrarily $\forall \av$ \;
$\langle \xi,\av \rangle \leftarrow \langle \xi_0,\pi^\epsilon(\xi_0) \rangle.$
\While{time left}{
Take action $a,$ observe reward $R$ and next state $\xi^\prime$
\begin{align*}
& Q^+(\xi_i,\av) \leftarrow R + \gamma \max_{\av^\prime} Q(\xi_i^\prime,\av^\prime) \qquad \qquad  \mathcal{O} (ndA^2)\\
& \delta \leftarrow Q^+(\xi_i) - Q(\xi_i) \\
& \thetav(\av) \leftarrow \thetav (\av) + \alpha \delta \phi (\xi_i)  \ \ \ \qquad \qquad \qquad \qquad \ \mathcal{O} (nd A)
\end{align*}
$\langle \xi^\prime,\av \rangle \leftarrow \langle \xi^\prime,\pi^\epsilon(\xi^\prime) \rangle$  $\qquad \qquad \qquad  \qquad \  \  \
 \mathcal{O} (nd A^2)$

}

Return $\pi$ greedy with respect to $Q$
\caption{\small Q-Learning with function approximation}
\end{algorithm}

{\bf Computational complexity of QLFA algorithm:} 
The computational complexity of main execution steps of the QLFA algorithm is indicated in Algorithm~2. We note that computing $Q(\xi_i,\av)$ has a complexity of $\mathcal{O}(ndA)$ (since it involves multiplication of two vectors of dimension $ndA$ as in eq.~\eqref{eqn:Qfromfeats}).
Thus, computing the policy $\max_a Q(\xi_i,a)$ has a complexity of $\mathcal{O}(ndA^2).$ 
We observe that the complexity of the algorithm is linear in the system dimension $n,$ as opposed to the exponential complexity of the value iteration and the original Q-learning algorithm. Moreover, the number of parameters to be stored and updated is $ndA.$ Thus, The QLFA algorithm is feasible even for high-dimensional systems.

In Section~\ref{sec:Sim_Res}, we conduct simulations to compare the performance of attack sequences generated by the QLFA algorithm with other attack sequences in high-dimensional systems. 

\subsection{Q-learning with Non-Linear Function Approximation}
Although the QLFA algorithm is computationally lightweight, its performance often requires a careful design of the feature set, which is non-trivial and challenging  \cite{Roy2013Tutorial}. 
Without the right set of features, the linear function approximators may not accurately represent the true value function. Alternately, one can use non-linear function approximators (NLFA) such as a neural network. The NLFA offer a richer class of function approximators that can map directly from the states to the value function without the need to explicitly specify the feature set. In particlar, we use the Q-learning with non-linear function approximation (Q-NLFA) algorithm \cite{DQN13} to solve the MDP \eqref{eqn:attacker_problem}. Simulation results for the Q-NLFA algorithm are presented in Section~\ref{sec:Sim_Res}-E, where we show that it outperforms the QLFA algorithm in high-dimensional systems.

Finally, note that in this section, we have used the \emph{value function approximation} method (i.e., we approximated the value function using linear/non-linear approximators). An alternative function approximation approach in high-dimensional/continuous systems is the \emph{policy gradient} method, which approximates the policy directly instead of the value function \cite{Sutton1999}. While policy gradient methods are well-suited for high-dimensional/continuous systems, they are known to exhibit high variance in estimating the gradient, which may adversely affect the learning.

\section{Simulation results}
\label{sec:Sim_Res}
We perform simulations on the voltage control system using PowerWorld, which is a high-fidelity power system simulator widely used in the industry \cite{PowerWorld}. 
All the simulations are performed based on the IEEE 9-bus system shown in Fig.~\ref{fig:9_bus}, in which buses $1$, $2$, and $3$ are the generator buses, buses $5$, $6$ and $8$ are the load buses (we use the default load values), and $5$, $7$, and $9$ are the pilot buses. The control matrix $\Bm$ is estimated using linear regression on the data traces of $\xv[t+1]-\xv[t]$ and $\uv[t]$ obtained in a PowerWorld simulation. Specifically, we applied a series of control inputs (i.e., changed the voltage of the generator bus) and observed the change in voltage of the pilot bus in Powerworld simulation, and used linear regression between the two data series to obtain $\Bm.$ Further, the process noise and observation noise are assumed to be Gaussian with standard deviation of $0.01$ and $0.02~$pu respectively. 

\begin{figure}[!t]
\centering
\includegraphics[width=0.45\textwidth,trim={0 7cm 0 0}]{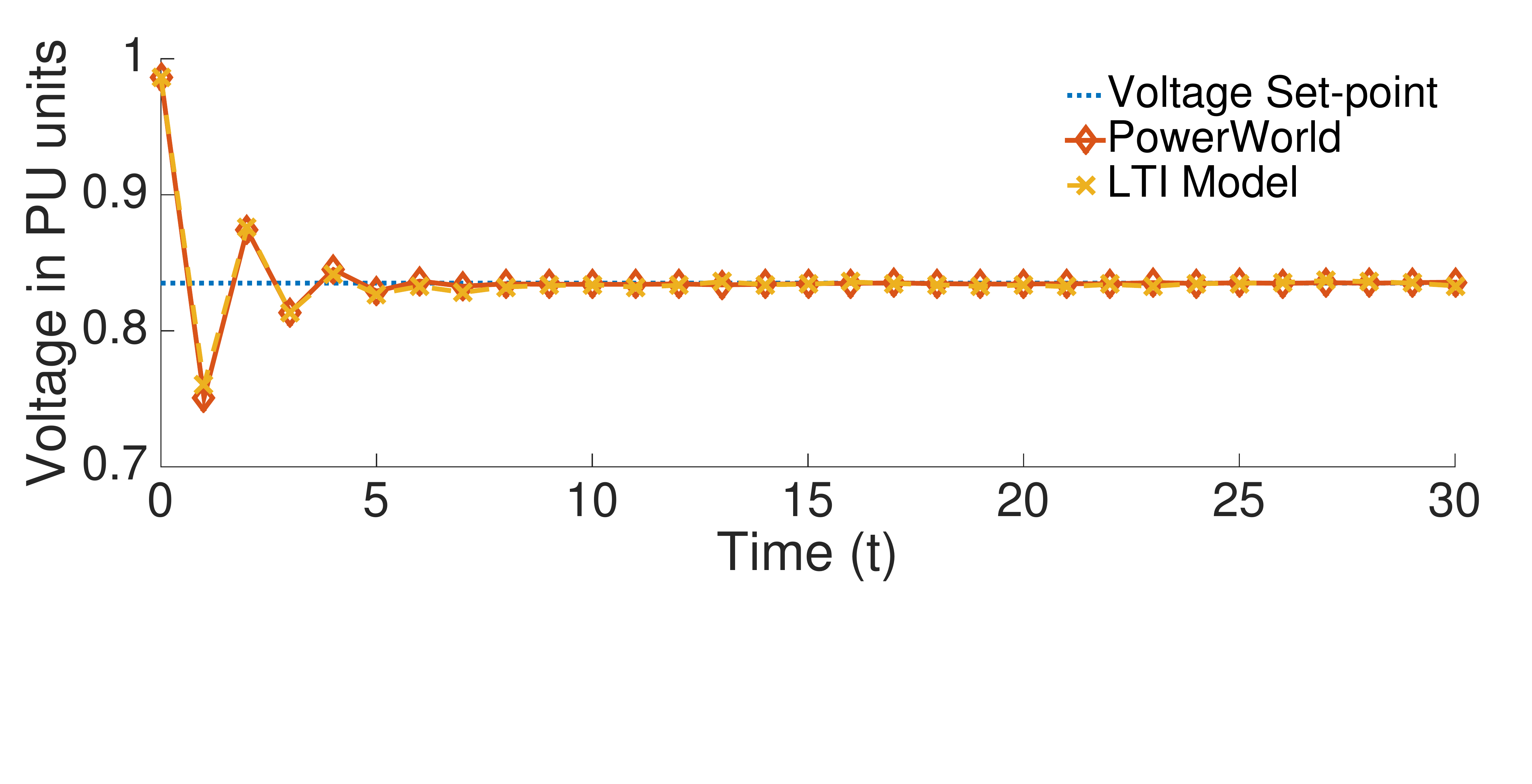}
\caption{Comparison between PowerWorld and LTI model.}
\label{fig:comaprsion_PW_LTI}
\end{figure}

First, we verify the accuracy of the LTI model in approximating the real-world voltage control system by examining the voltage at pilot bus $5$. In our simulations, the voltage controller aims to adjust the voltage of this bus from an initial voltage of $1$~pu to a setpoint ($\xv_0$) of $0.835$~pu (base voltage of $230$~kV) by applying the control by applying the control $\uv[t] = \Bm^{-1} (\xv_0 - \widehat{\xv}[t])$ in both Powerworld and the LTI model. Fig.~\ref{fig:comaprsion_PW_LTI} plots the bus voltage from $t = 1$ to $t =30$ obtained from the PowerWorld simulations, as well as the voltage values obtained from the LTI model. To average the effect of random measurement noise, we repeat the experiment $100$ times, and take the mean value. The two curves match
well in this figure, thus verifying the accuracy of the proposed LTI model. 

We also include a simulation result to show the validity of the approximation that the attacker can accurately track the KF's estimation error (see bullet point~1 in Section 
~V). We consider the IEEE-9 bus system with similar settings as above and under a ramp attack sequence. We set the slope of the ramp to $0.01$ pu. For simplicity, we do not consider attack detection (since our objective is to show the attacker's estimate of the system state under an attack sequence that bypasses detection). Denote the defender's and the attacker's estimates of the system state 
by $\widehat{\xv} [t] $ and $\widehat{\xv}_a [t]$ respectively, which are computed as
\begin{align*}
\widehat{\xv} [t+1] & \!=\! \Am \widehat{\xv}[t]\! +\! \Bm \uv[t] \! +\! \Km (\yv_f [t+1]\! -\! \Cm (\Am \widehat{\xv}[t]\! +\! \Bm \uv[t])),  \\
\widehat{\xv}_a [t+1] & \!=\! \Am \widehat{\xv}_a[t]\! +\! \Bm \uv[t] \! +\! \Km (\yv [t+1]\! -\! \Cm (\Am \widehat{\xv}[t]\! +\! \Bm \uv[t])).
\end{align*}
Here $\yv_f[t] = \yv[t] - i[t] \deltav[t]$ and $\uv[t] = \alpha \LB \xv_0 - \widehat{\xv}[t]\RB.$
Note that the attacker's estimate is computed using $\yv [t+1]$ (i.e., the true sensor measurements).  

\begin{figure}[!t]
  \subfigure[]
  {
    \includegraphics[width=0.47\textwidth,trim={0 7cm 0 0}]{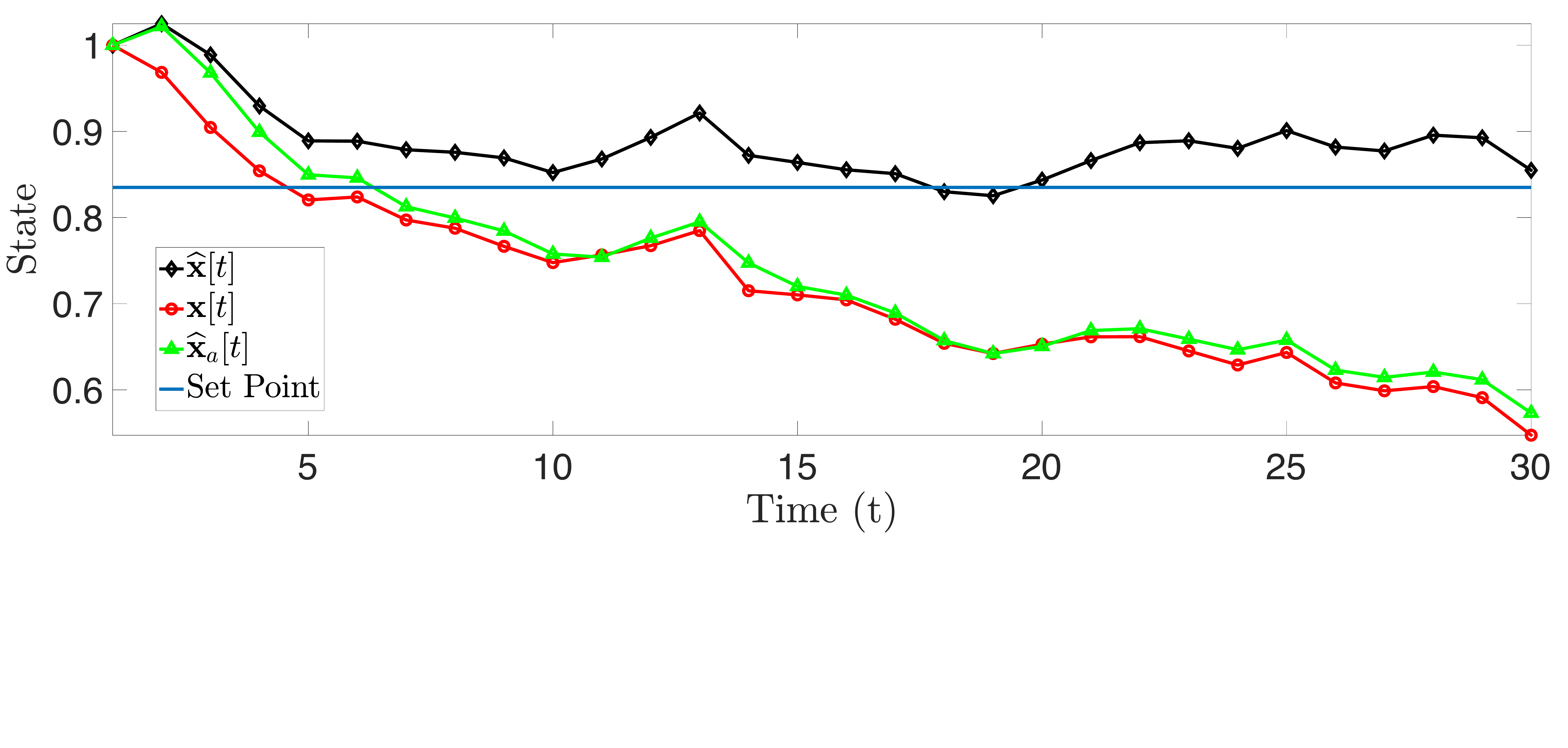}
  }
  \subfigure[]
  {
    \includegraphics[width=0.47\textwidth,trim={0 7cm 0 0}]{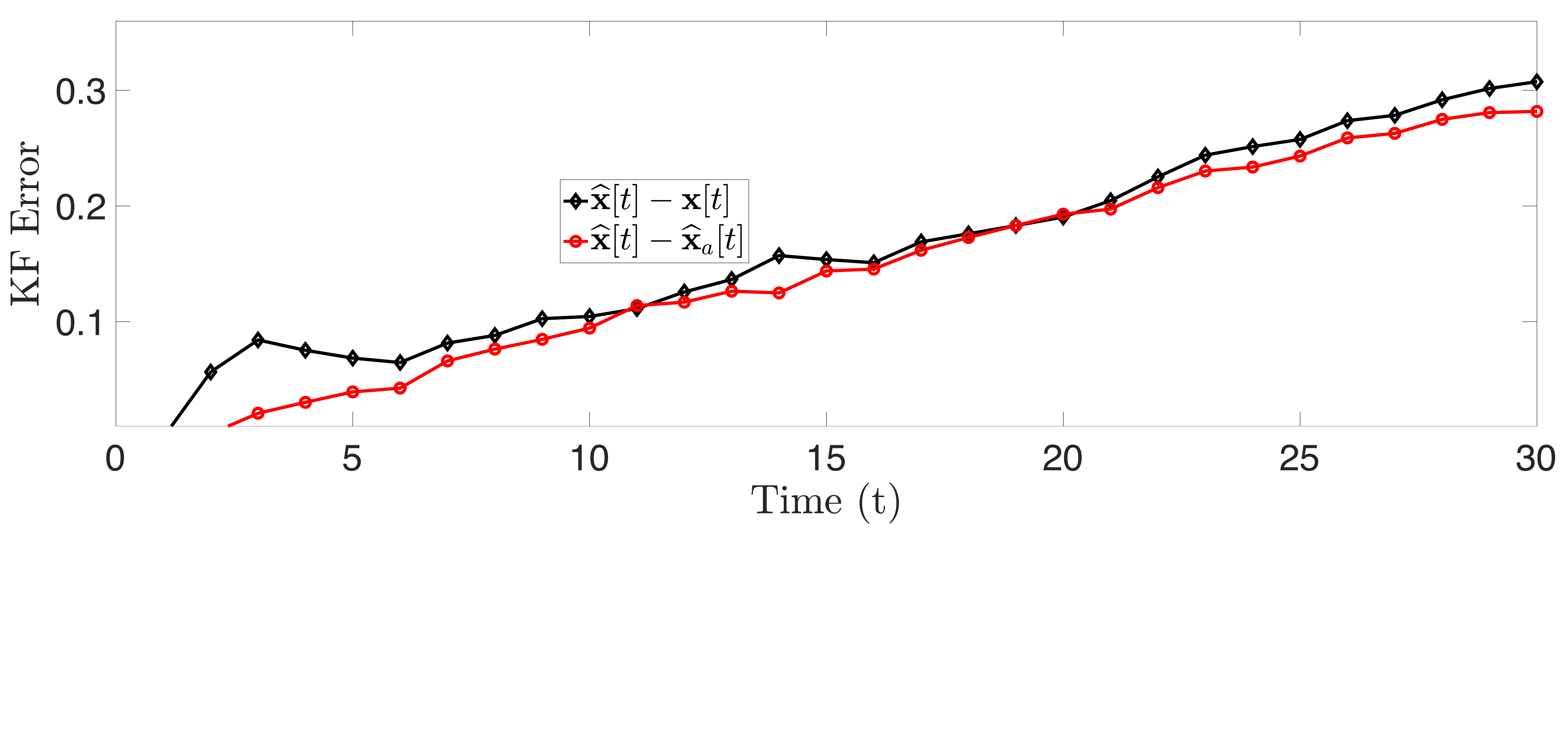}
  }
  \vspace{-1em}
\caption{IEEE-9 bus system under ramp attack. (a) Evolution of system state and the estimate. (b) Evolution of the KF estimation error }
\label{fig:Attacker_Estimate}
\vspace{-1em}
\end{figure}

The result of the simulation is plotted in Fig.~\ref{fig:Attacker_Estimate} (a). It can be seen that while the defender (i.e., the system operator) has an erroneous estimate of the system state due to the FDI attack, the attacker can track the system state fairly accurately based on the true sensor measurements. We also plot the KF estimation error (i.e., $ \widehat{\xv}[t] - \xv[t]$), and the attacker's estimate of the KF error (i.e., $\widehat{\xv}[t] - \widehat{\xv}_a[t] $) in Fig.~\ref{fig:Attacker_Estimate} (b), which shows a close match between the two curves. Hence, in our analysis, we assume that the attacker can track the Kalman filter's estimation error $\ev[t] $ accurately.

\subsection{Attacks Against Voltage Control Using MDP and Value Iteration Algorithm}
\begin{figure}[!t]
  \subfigure[]
  {
    \includegraphics[width=0.47\textwidth,trim={0 5cm 0 0}]{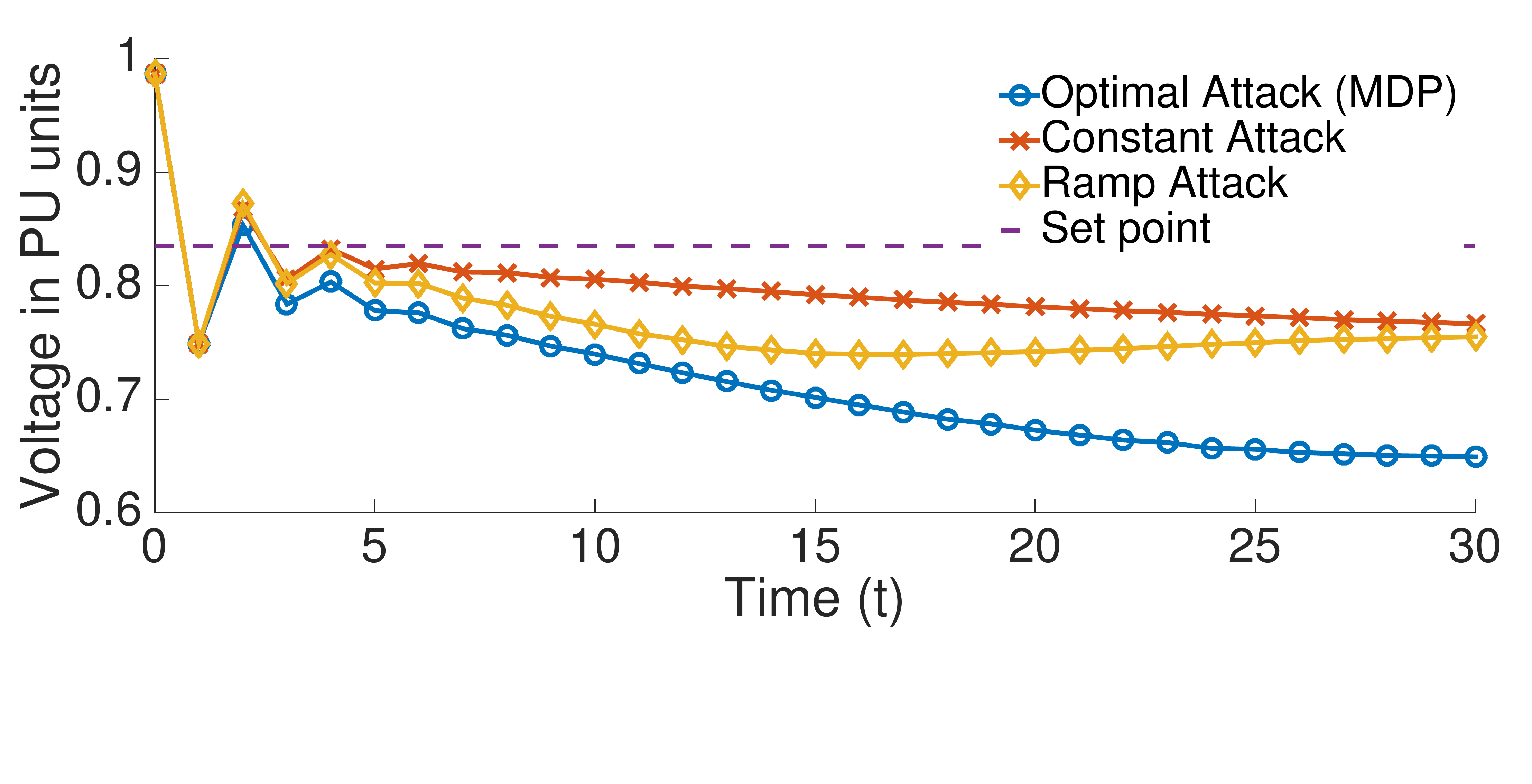}
  }
  \subfigure[]
  {
    \includegraphics[width=0.47\textwidth,trim={0 5cm 0 0}]{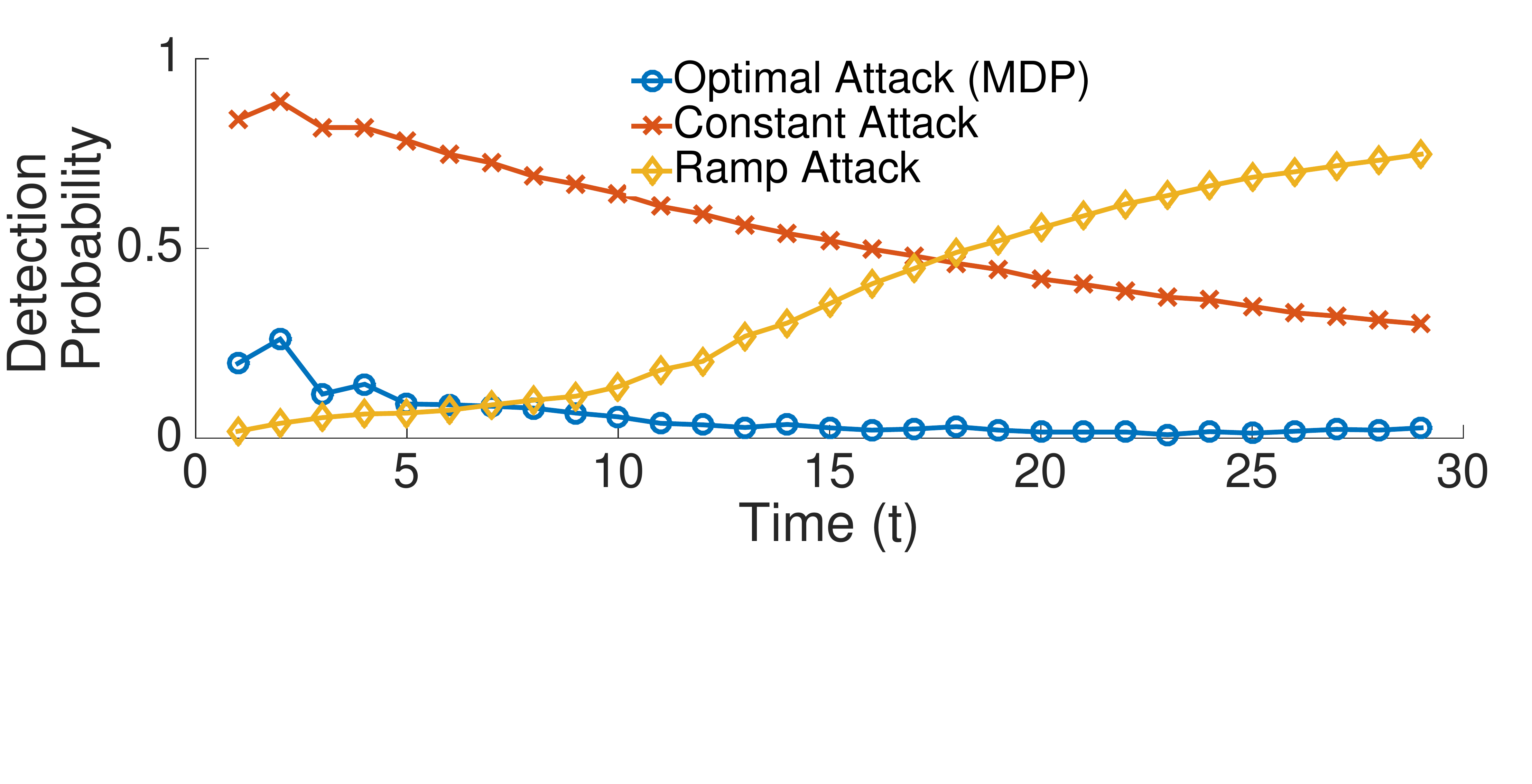}
  }
\caption{Attack against voltage control system. (a) Pilot bus voltage (Bus 5) under different attack sequences. (b) Attack Detection probability for different attacks.}
\label{fig:voltage_state}
\end{figure}

Next, we simulate the impact of the proposed attacks on the voltage control system while considering the $\chi^2$ attack detection. We assume that the attacker has access to the voltage sensor of bus $5,$ and injects false measurements to mislead the controller. We compute the optimal attack sequence 
based on the LTI model using the value iteration method implemented in MATLAB. 
Recall that in this example, the state of the MDP is the voltage of bus~5 and the attack sequence is the FDI attack on the voltage sensor measurement of bus~5. We restrict $\av[t]$ between $0-0.2$~pu discretized with an interval of $0.02~$pu. To evaluate the attack impact, we run Monte Carlo simulations using the PowerWorld simulator by injecting the derived optimal attack into the voltage measurements, and implementing the control based on the corresponding state estimate.
Fig.~\ref{fig:voltage_state} (top figure) shows the pilot bus voltage (bus 5) for different attack sequences with $\eta = 5$ and perfect attack mitigation. It can be observed that the pilot bus voltage deviates from the setpoint of $0.835$~pu, and the largest voltage deviation is seen under the optimal attack. In particular, over an attack duration of $30$ time slots, we observe that bus $5$ voltage deviates to $0.65$~pu under the optimal attack, which is about $0.2$~pu from the setpoint.

\begin{figure}[!t]
  \subfigure[]
  {
    \includegraphics[width=0.225\textwidth]{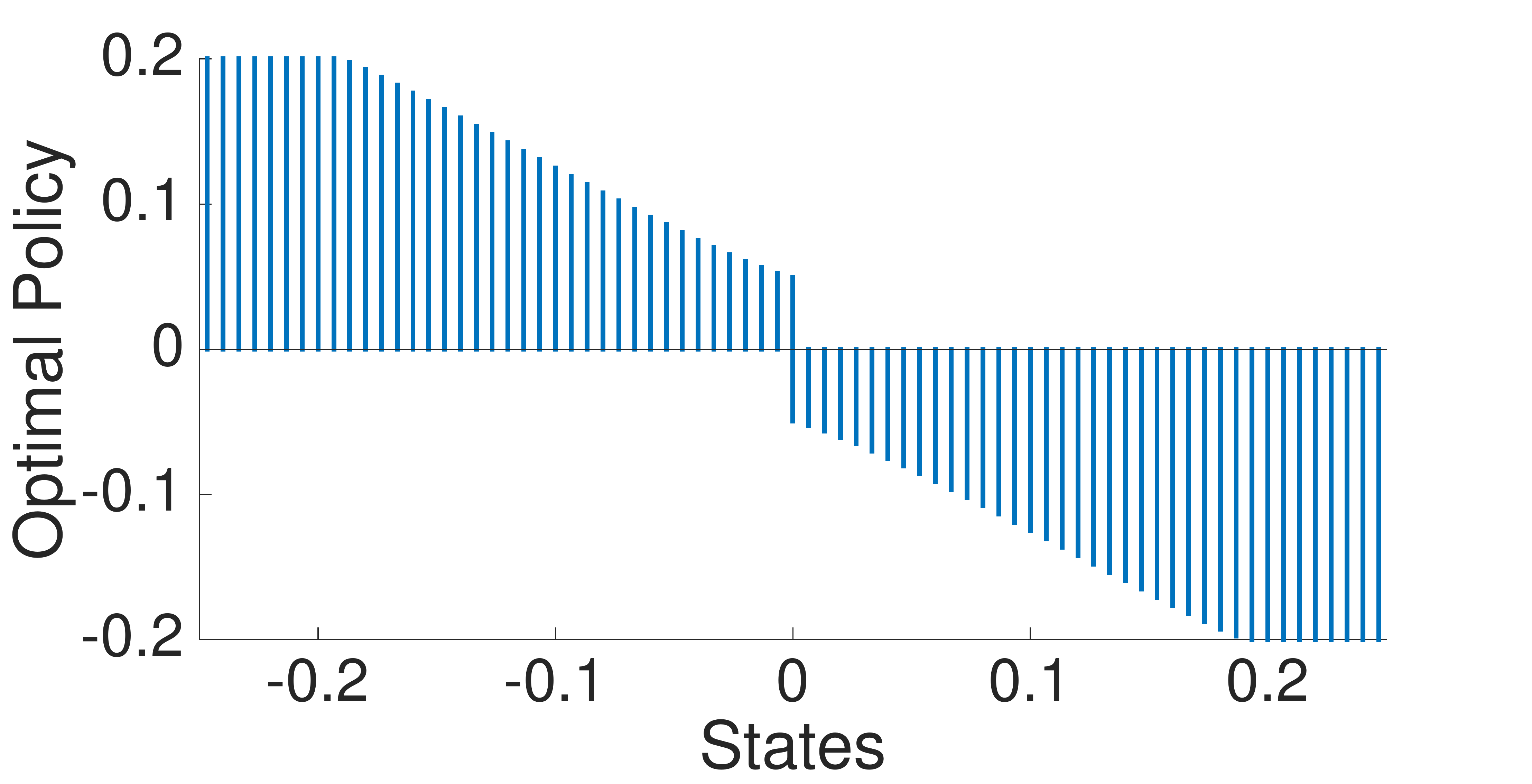}
  }
  \subfigure[]
  {
    \includegraphics[width=0.225\textwidth]{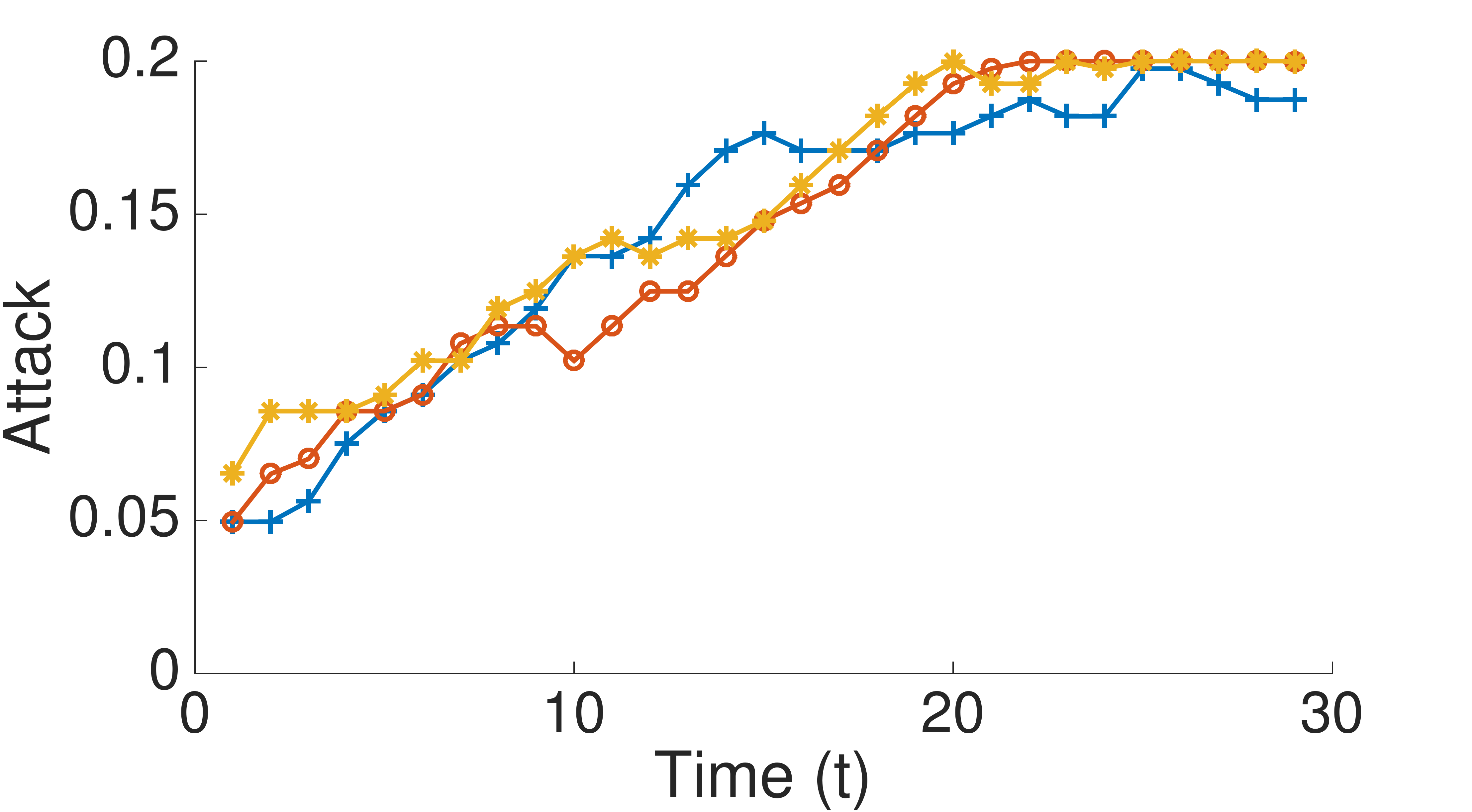}
  }
\caption{(a) Optimal policy for different system states computed by value iteration. (b) Optimal attack sequence for 3 Monte Carlo simulation instantiations.}
\label{fig:optimal_policy}
\end{figure}

Fig.~\ref{fig:voltage_state} (bottom figure) shows the attack detection probability under these attacks at different time instants. We also plot the optimal policy computed by the value iteration algorithm (Algorithm~1) in Fig.~\ref{fig:optimal_policy}a, and the optimal attack sequence for three Monte Carlo instantiations in Fig.~\ref{fig:optimal_policy}b. We observe that the attack detection probability for a naive attack sequence such as the ramp attack increases with time, which results in nullifying its impact due to attack mitigation. However, the optimal attack is crafted in a way such that the detection probability decreases over time. Consequently, the optimal attack causes a significant deviation of the pilot bus voltage from its setpoint.

\begin{figure}[!t]
  \subfigure[]
  {
    \includegraphics[width=0.45\textwidth]{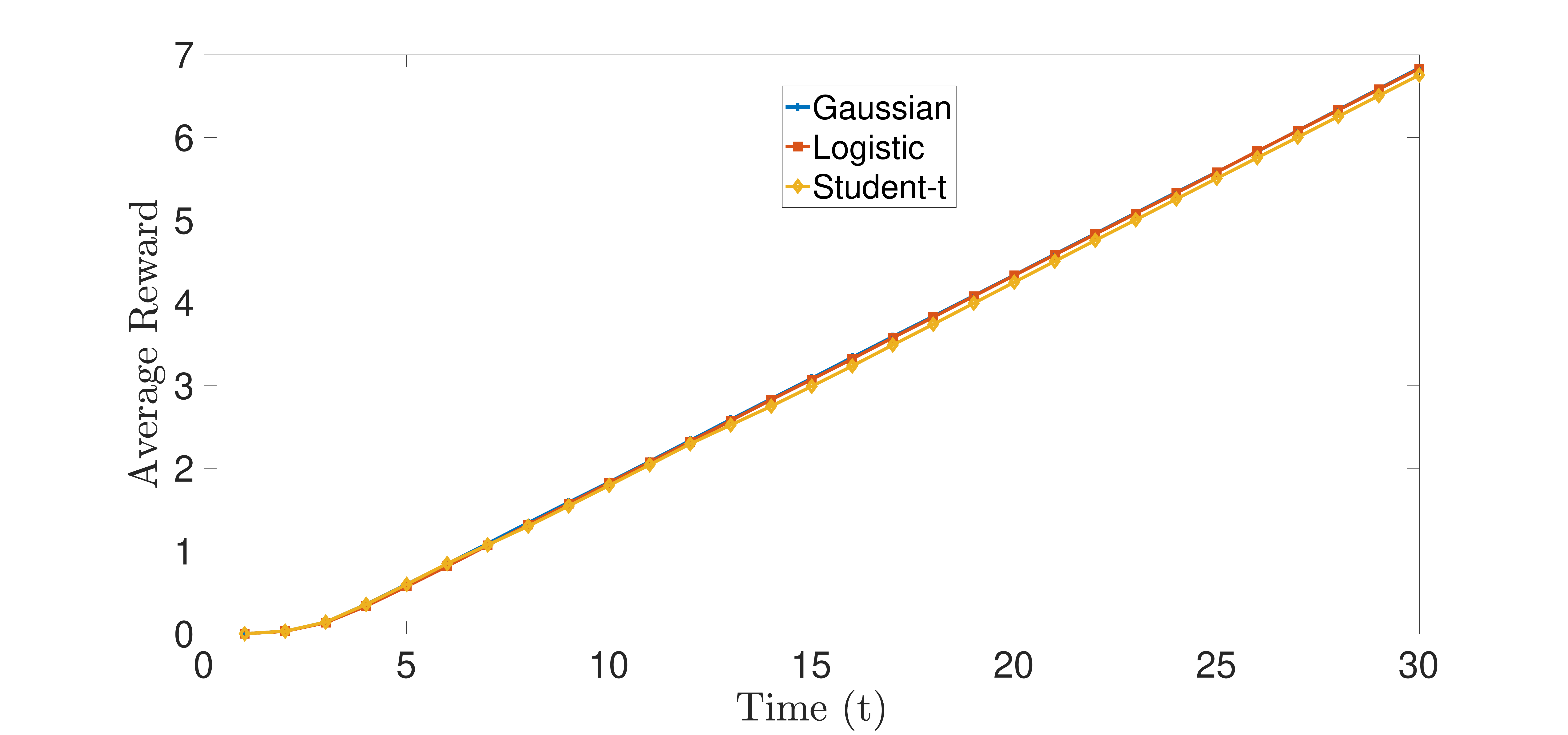}
  }
  \subfigure[]
  {
    \includegraphics[width=0.45\textwidth]{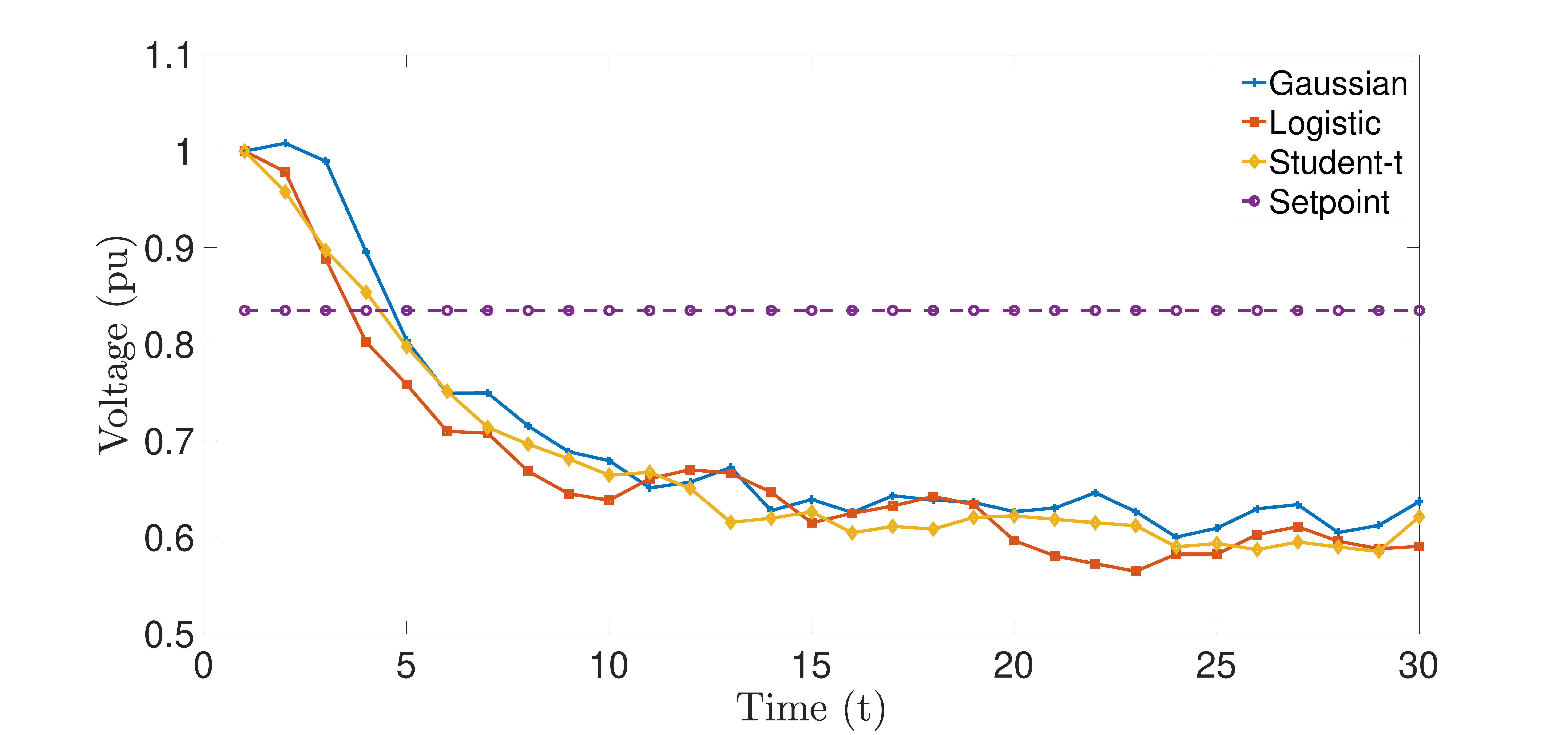}
  }
\caption{(a) Average reward of the MDP policy, (b) pilot bus voltages under non-Gaussian noise settings.}
\label{fig:nongaussian}
\end{figure}

{\bf Algorithm Performance under Non-Gaussian Noises:}  We also perform simulations to examine the algorithm performance under non-Gaussian noise settings. Recent work based on real PMU noise data shows that its statistical distribution can be modeled by non-Gaussian distributions such as the Student-t or Laplace distributions  \cite{Wang2018}. Accordingly, we examine the performance of the policy derived using the MDP algorithm in a power system whose measurement noises follow a non-Gaussian distribution. Specifically, we derive the policy using the MDP approach (which assumes Gaussian noise). However, at the policy evaluation stage, we evaluate using two non-Gaussian noise distributions, i.e., Logistic and Student-t distributions and compare the time average cost function (time average Kalman filter estimation error). For fair comparison, we set the noise variance to be the same as that of the Gaussian case (0.02~pu). The results are shown in Fig.~\ref{fig:nongaussian}~(a), in which we plot the time average cost function. It can be seen that the performance of the MDP algorithm with non-Gaussian noise is nearly identical to the Gaussian noise setting. We also plot the evolution of the pilot bus voltage in the presence of attack based on the MDP policy in Fig.~\ref{fig:nongaussian}~(b). The performance degradation is nearly identical in all the three cases. Hence, we conclude that the MDP algorithm is nearly optimal for real-world PMU sensor measurement noises. 

\subsection{Quantifying Cost of False Positives and Missed Detections}
\label{sec:quant_MD_FP}
\begin{figure*}[!t]
  \subfigure[$\sigma_{\text{mit}} = 0$]
  {
    \includegraphics[width=0.45\textwidth]{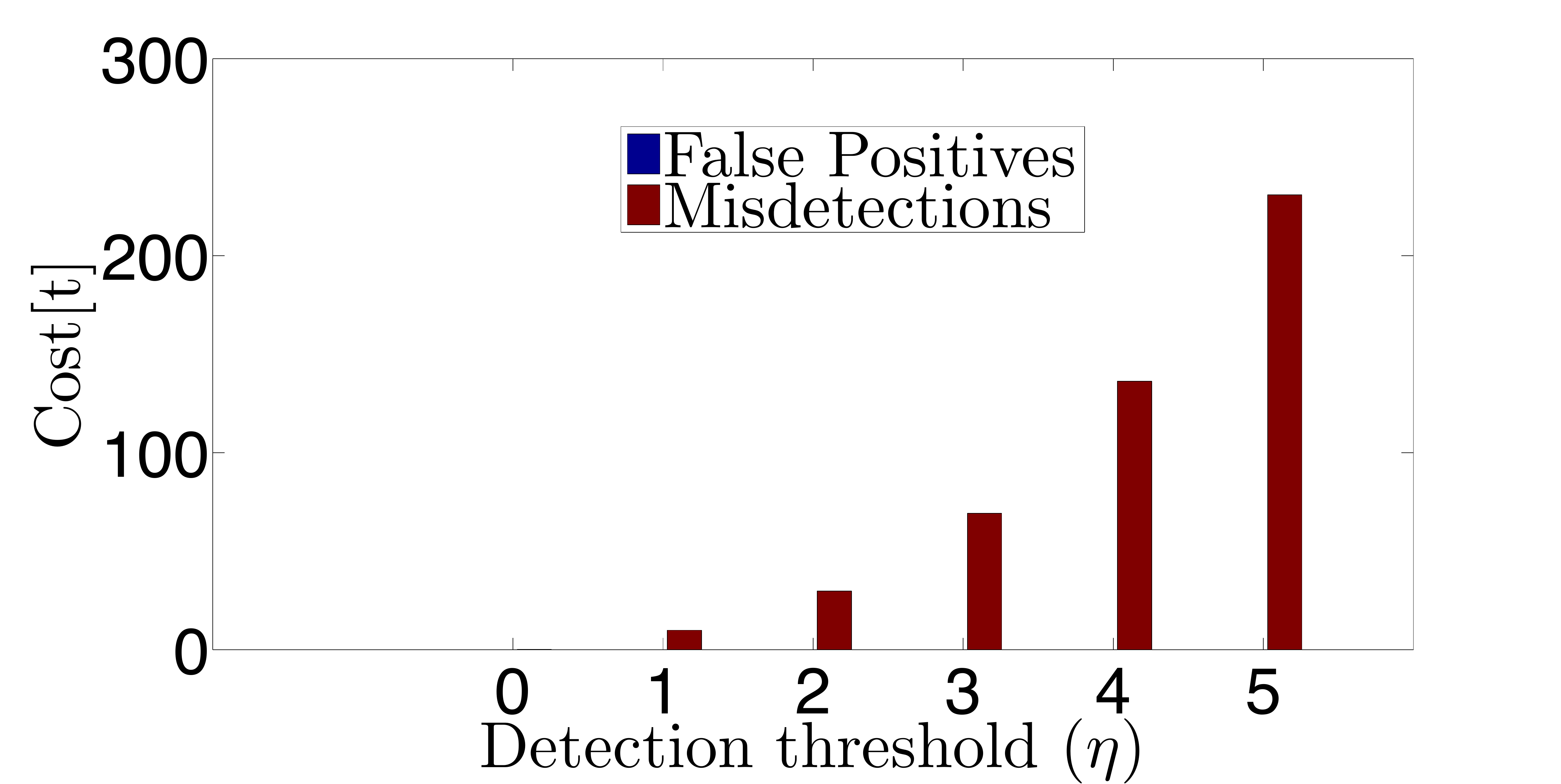}
  }
  \subfigure[$\sigma_{\text{mit}} = 5$]
  {
    \includegraphics[width=0.45\textwidth]{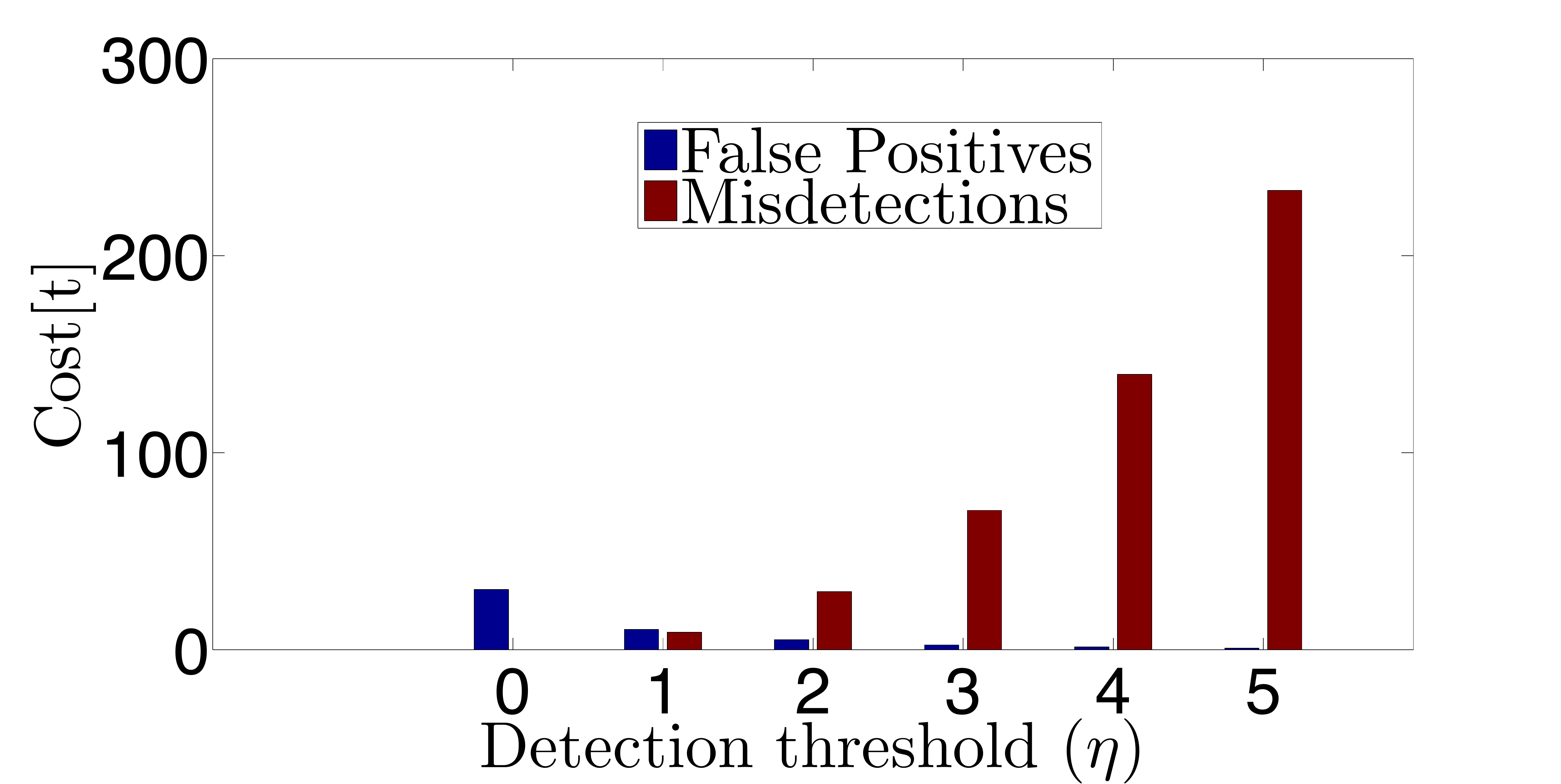}
  }
  \subfigure[$\sigma_{\text{mit}} = 10$]
  {
    \includegraphics[width=0.45\textwidth]{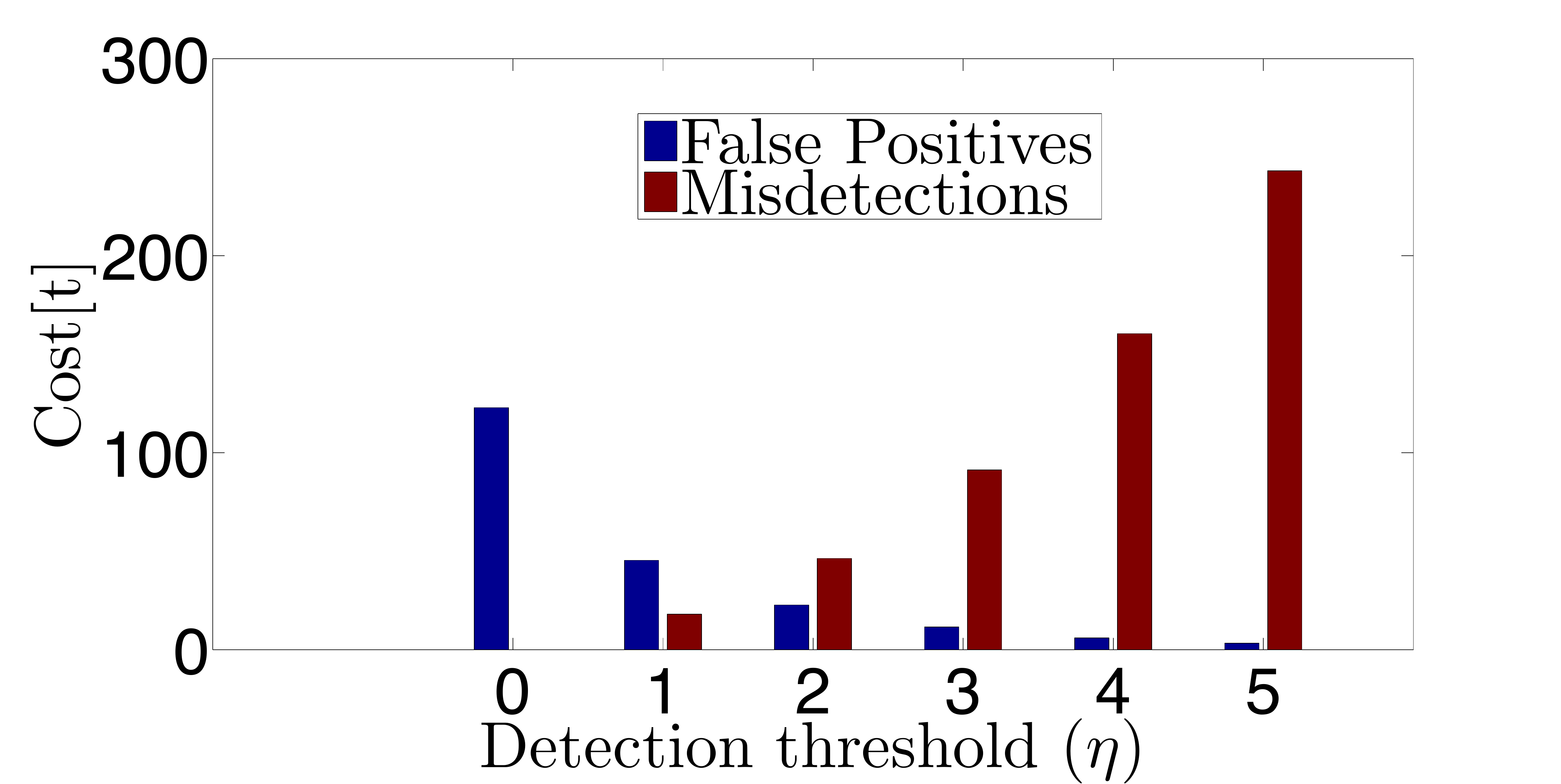}
  }
  \subfigure[$\sigma_{\text{mit}} = 15$]
  {
    \includegraphics[width=0.45\textwidth]{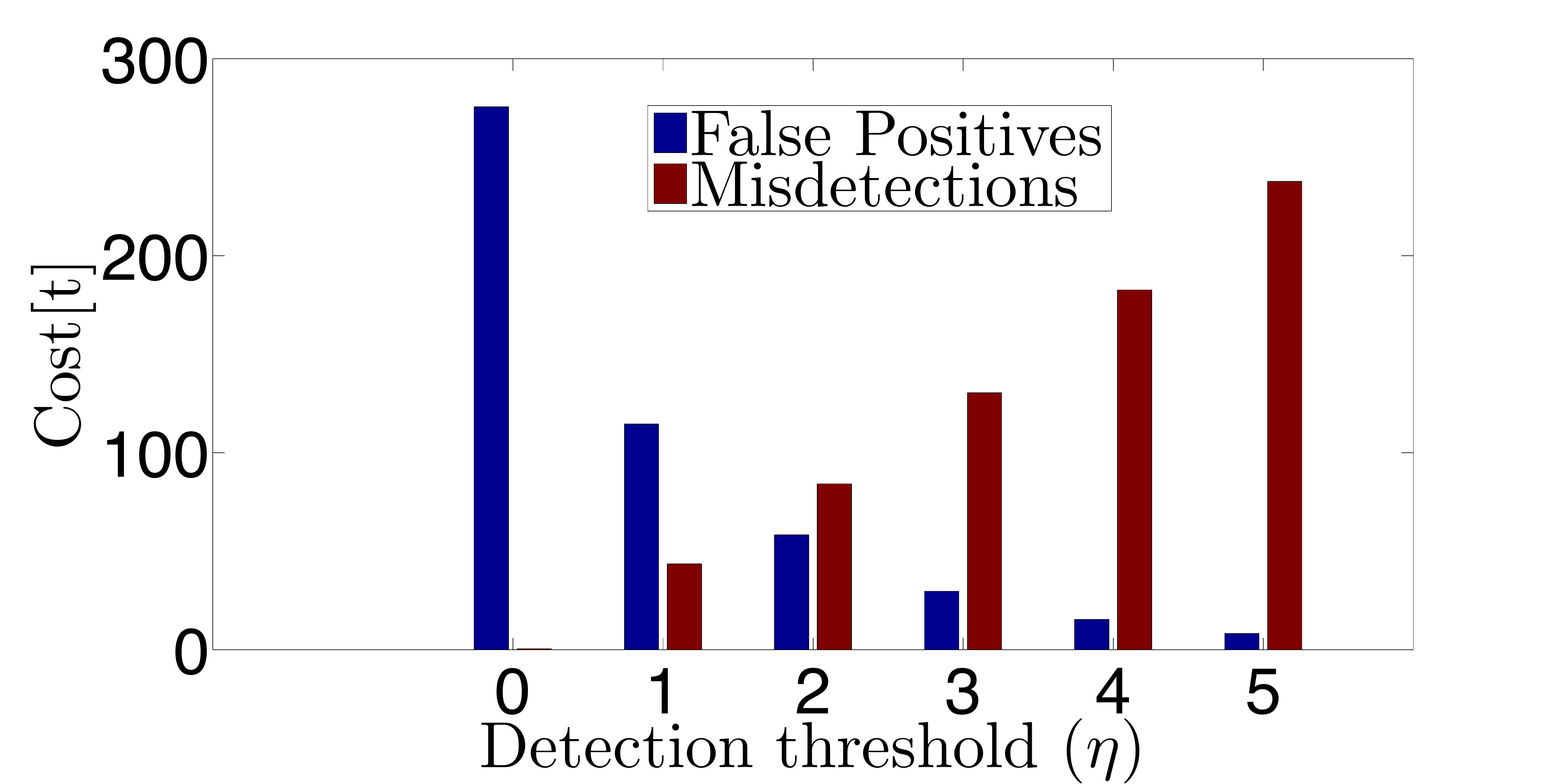}
  }
  \vspace{-1em}
  \caption{Cost of FPs and MDs for different attack detection thresholds and standard deviation of the attack mitigation signal $\sigma_{\text{mit}}$.}
  \label{fig:Thold}
\end{figure*}

In this subsection, we perform simulations to quantify the cost of FPs and MDs following the approach in Section~\ref{sec:MD_FP}. 
We consider a general LTI model described by \eqref{eqn:process} and \eqref{eqn:Obs} with $n = 1,$ $\Am$ = $1$, $\Cm$ = $1$, $\Qm$ = $1$, $\Rm$ = $10$  and adopt a \emph{practical attack mitigation approach} under which the attack mitigation signal is given by $\deltav_d[t] = \av[t]+\bv[t],$ where $\bv[t]$ is the error component in the mitigation signal. We generate $\bv[t]$ as a Gaussian distributed random variable with a standard deviation of $\sigma_{\text{mit}}$ units. Fig.~\ref{fig:Thold} shows the cost of FPs and MDs for different detection thresholds $\eta$ and standard deviations of the attack mitigation signal $\sigma_{\text{mit}}$. 

From these plots, we observe that as the attack detection threshold $\eta$ is increased, the cost of FP decreases and the cost of MD increases. This result is intuitive -- a low detection threshold detects most attacks but also leads to a large number of FPs. Thus, the wrongly triggered mitigations will result in a high FP cost. On the other hand, a high detection threshold yields a low number of FPs, but also increases the number of MDs. The figures show a basic tradeoff between FPs and MDs, quantified in terms of the cost function. 

We also observe that these costs depend on the accuracy of the attack mitigation signal. For instance, when the accuracy is high (e.g.,  $\sigma_{\text{mit}} = 0, 5),$ the cost of FP is very low,
even for a low detection threshold. Thus, in this scenario, the system operator can choose a low detection threshold and obtain good overall system performance. However, when the accuracy of the mitigation signal is low (e.g., $\sigma_{\text{mit}} = 15),$ the cost of FP is high for a low detection threshold. For instance, in Fig.~\ref{fig:Thold}(d), the cost of FP for $\eta = 0$ is greater than the cost of MD for $\eta = 5.$ In this scenario, the system operator must choose a high detection threshold to obtain an acceptable level of system performance. 
Thus, our result helps the system operator select an appropriate threshold to
balance the costs of FP and MD, depending on the accuracy of the mitigation signal.

Lastly, we note that for $\sigma_{\text{mit}} = 0$ (i.e., perfect mitigation), the cost of 
FP is zero for all detection thresholds. Under perfect mitigation, even 
if an FP event occurs, the controller can accurately estimate that the attack magnitude is zero (i.e., no attack). Thus, in this specific case, wrongly triggered mitigations do not increase the cost of FP.
We also note that for $\eta = 0,$ there are no MDs. Hence, the cost of MD in this case is nearly zero.

\subsection{Comparison of Value Iteration, Q-Learning and QLFA Algorithms }
\begin{figure}[!t]
  \subfigure[]{
    \includegraphics[width=0.45\textwidth]{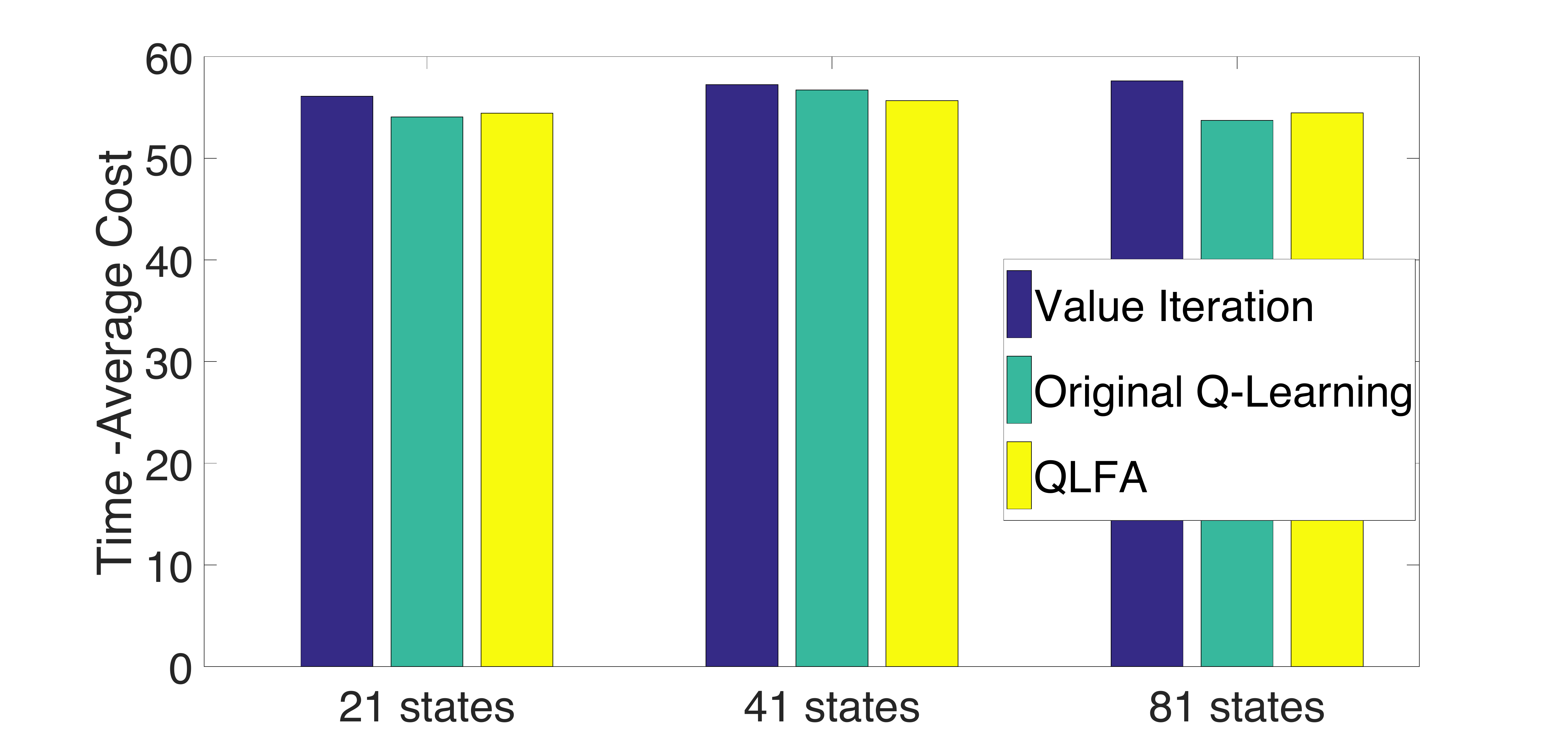}
  }
  \subfigure[]{
    \includegraphics[width=0.45\textwidth]{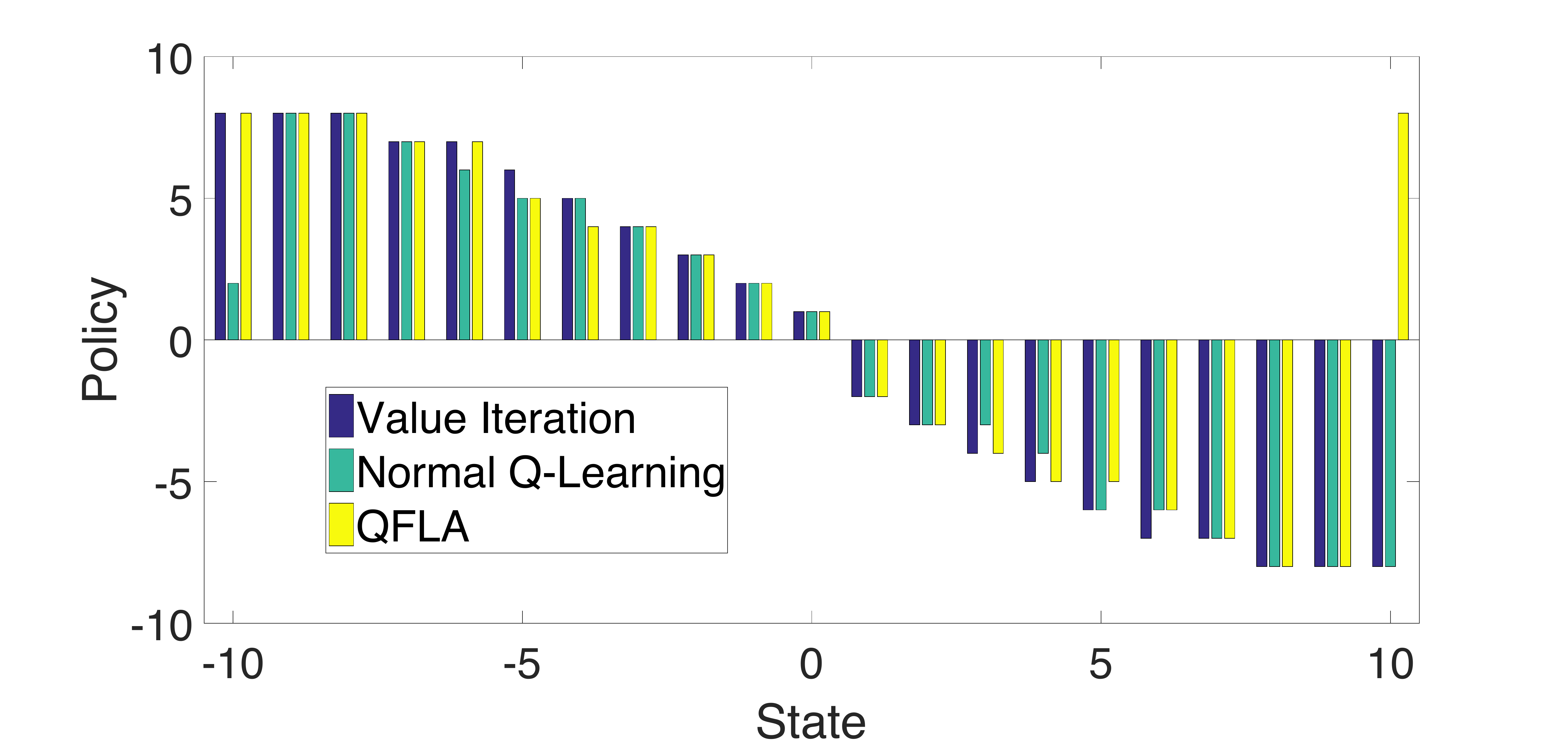}
  }
  \vspace{-1em}
  \caption{(a) Time average reward under policies from value iteration, naive QL and QLFA algorithms in 1-D system. (b) Policies from the respective algorithms for $d = 21$ states.}
  \label{fig:QLFA-results}
\end{figure}

For the LTI system described in previous subsection, we compare the performance of value iteration, Q-learning (eq.~\eqref{eqn:Q-Learn}) and QLFA algorithms in a $1$-dimensional system. For QLFA, we use FSR with binary encoding as the basis function. We first compute the policy $\pi(\ev)$ using each of the three algorithms and then evaluate the cost function using the corresponding policies. 
We repeat the simulations for different numbers of discretization levels $d = \{21,41,81\}.$
The results are shown in Fig.~\ref{fig:QLFA-results}a, which show that the performance of the original Q-learning algorithm and the QLFA algorithm is close to that of the value iteration algorithm. We also plot the policies of the three algorithms in Fig.~\ref{fig:QLFA-results}b, which are similar.

\subsection{Simulations for High-Dimensional Systems Using Q-Learning with Function Approximation}
We report simulations in high-dimensional systems using the QLFA (with FSR with binary encoding as the basis function) and Q-NLFA algorithms proposed in Section~\ref{sec:Q_learn}. We use two bus systems (i) a moderately-sized IEEE-39 bus system which has $10$ generator buses (bus $30-39$) and $10$ pilot buses and (ii) a large bus system, i.e., the IEEE-118 bus system which has $30$ pilot buses (we only assign the non-generator buses as pilot buses). The state of the system is the pilot bus voltages and the actions are the FDI attack injected at the pilot bus voltage sensor measurements. We estimate the control matrix $\Bm$ for both the bus systems using linear regression on data traces obtained in a PowerWorld simulation.

\begin{figure}[!t]
  \subfigure[]
  {
    \includegraphics[width=0.45\textwidth]{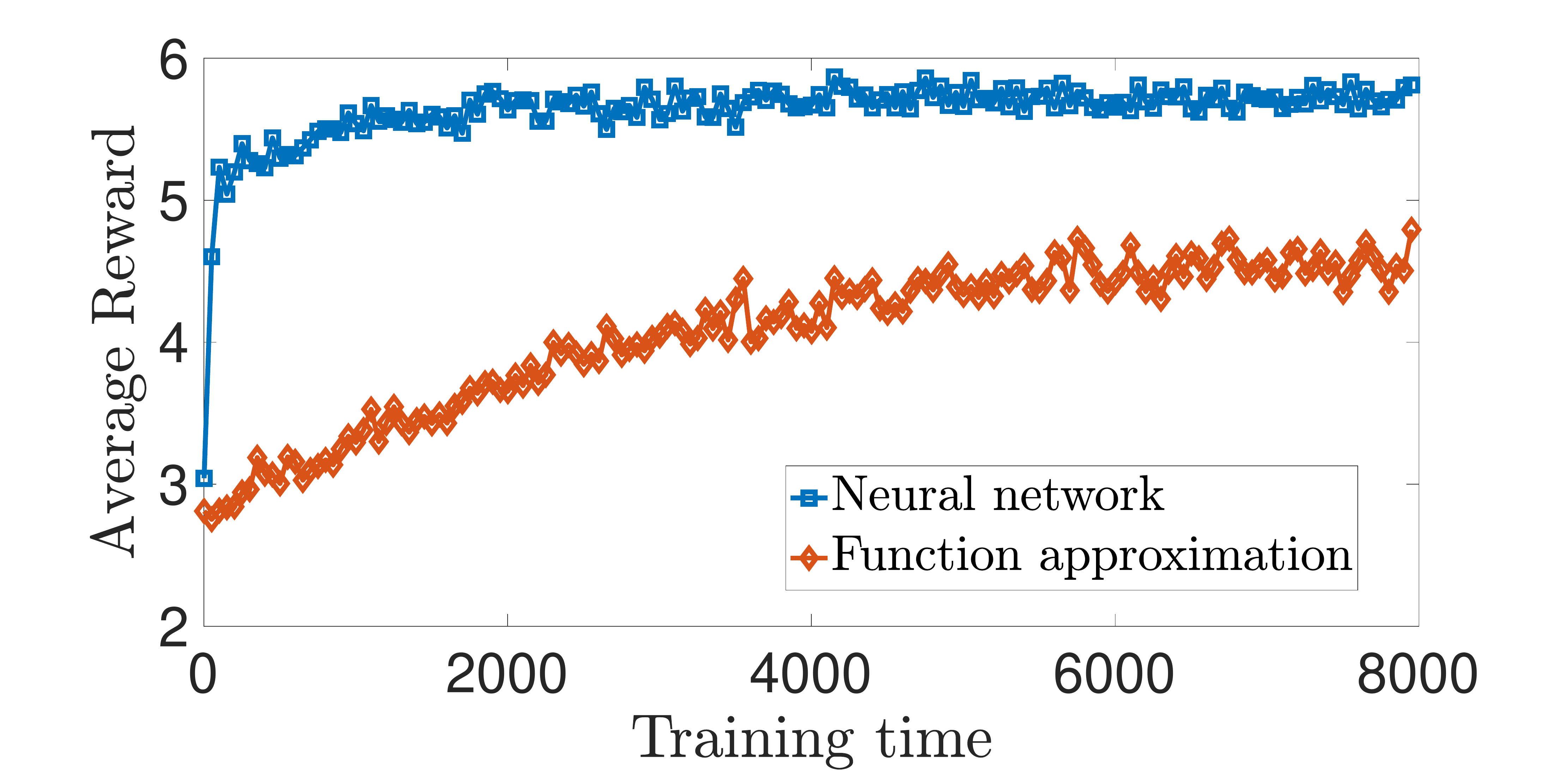}
  }
  \subfigure[]
  {
    \includegraphics[width=0.45\textwidth]{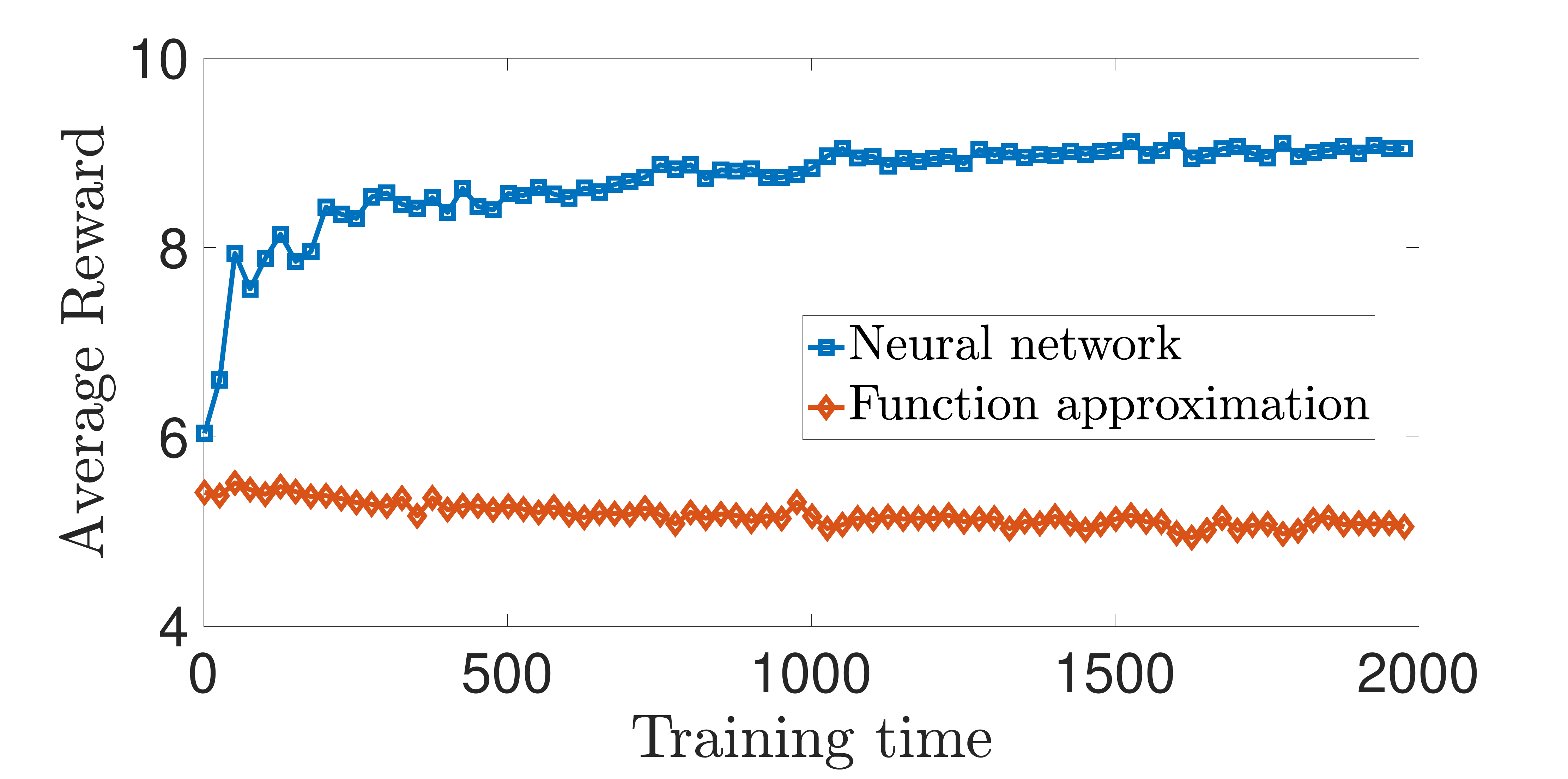}
  }
\caption{Comparison of QLFA and Q-NLFA algorithms for different training times (a) IEEE-39 bus system (with 10 pilot buses), (b) IEEE-118 bus system (with 30 pilot buses).}
\label{fig:DQN_sims}
\end{figure}

First, we show the average reward of the QLFA and Q-NFLA algorithms as a function of the training time in Fig.~\ref{fig:DQN_sims} for both the bus systems. To derive the policy for the QLFA algorithm, we compute the weight vector ($\thetav(\av)$) for different training times in a MATLAB simulation. To derive the policy for the Q-NLFA algorithm, we use the TensorFlow library \cite{TensorFlow}. In particular, we construct a neural network with $20$ nodes per layer and $5$ hidden layers and employ the rectified linear unit (ReLU) activation function. We set a batch size of $200$ for deep Q-learning's \emph{experience replay} step \cite{DQN13}.

After obtaining the policy in each case, we run test trajectories with the corresponding policies, and obtain the average reward (cumulative Kalman filter estimation error) over $50$ time slots. It can be observed from Fig.~\ref{fig:DQN_sims} that the average rewards of the QLFA and the Q-NFLA algorithms increase with the training time, implying that the algorithms learn policies that yield a high reward. Moreover, we observe that for the IEEE-39 bus system, the Q-NLFA algorithm converges to a policy that yields high reward much faster than the QLFA algorithm. For a larger system such as the IEEE-118 bus, the Q-NLFA algorithm significantly outperforms the QLFA algorithm, and the QLFA algorithm is suboptimal. Thus, we conclude that Q-NLFA algorithm should be used for high-dimensional systems (with more than 10 pilot buses).

\begin{figure}[!t]
  \subfigure[]
  {
    \includegraphics[width=0.47\textwidth]{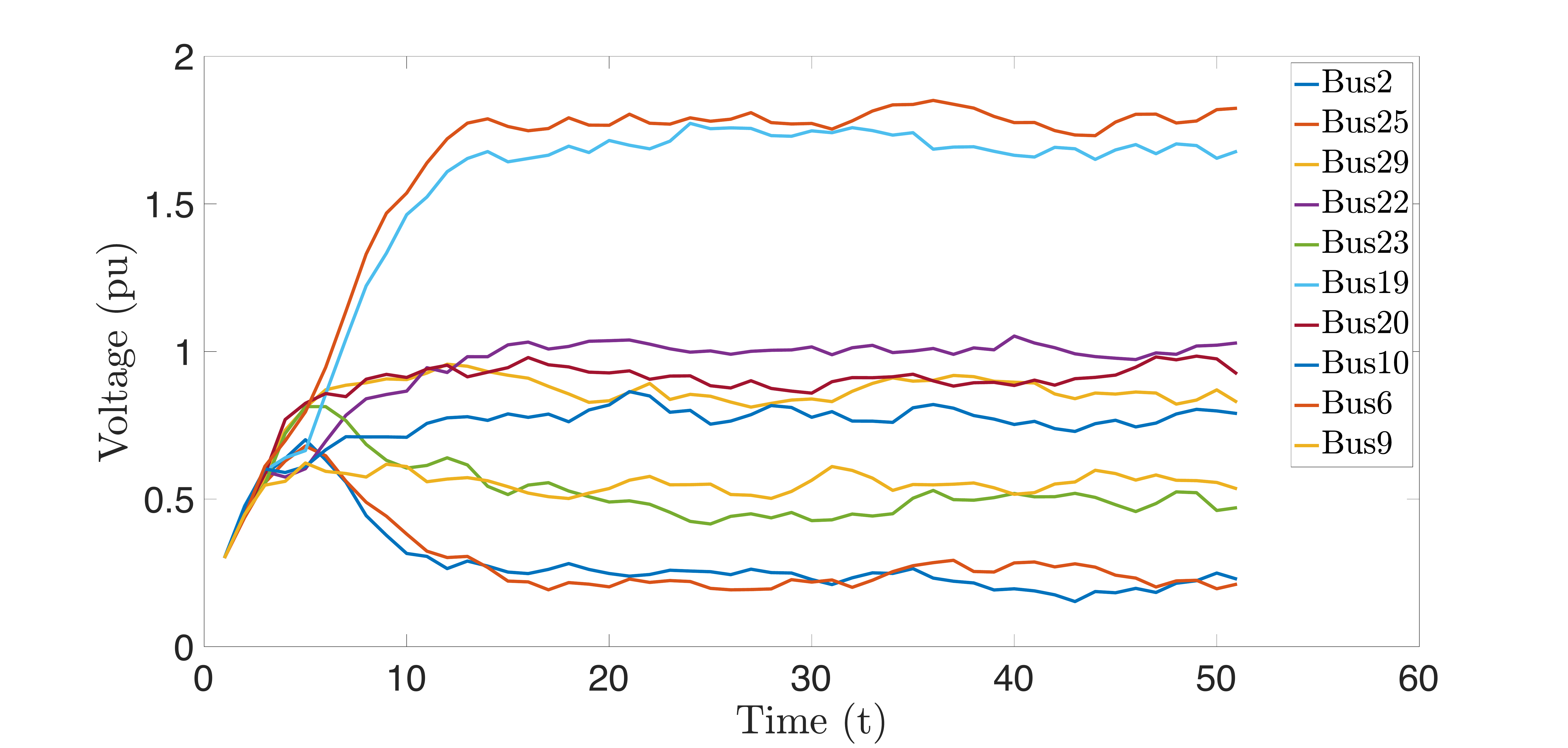}
  }
  \subfigure[]
  {
    \includegraphics[width=0.47\textwidth]{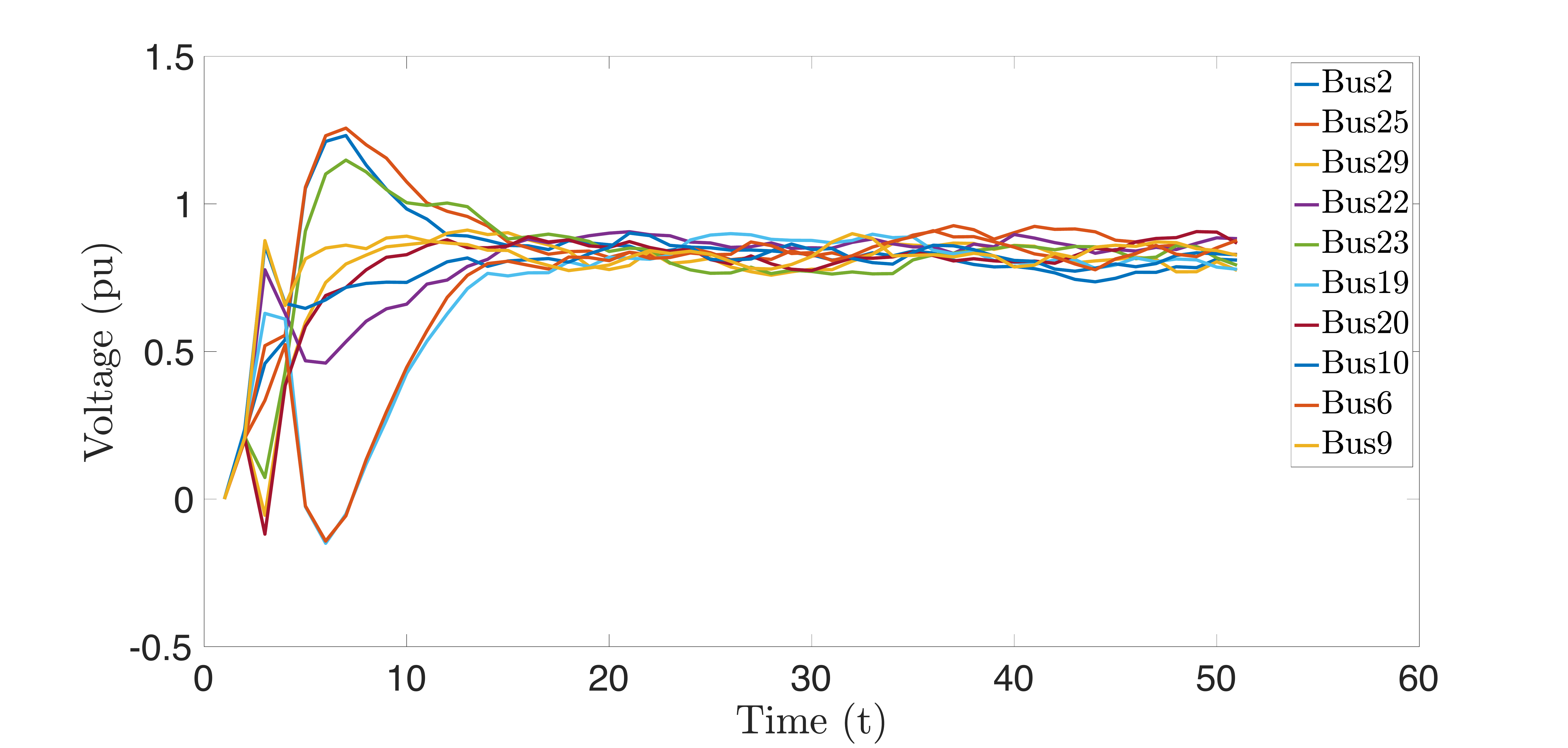}
  }
  \vspace{-1em}
\caption{Attack against voltage control system for the IEEE-39 bus. (a) Pilot bus voltages under QLFA attacks. (b) Estimates of the pilot bus voltages.}
\label{fig:IEEE39_voltage_state}
\vspace{-1em}
\end{figure}

Next, we use the policy obtained from the Q-NLFA algorithm as the attack vector in voltage control loop of the IEEE-39 bus system in a PowerWorld simulation. In these simulations, the controller adjusts the voltage of the pilot buses from an initial voltage of $0.3~$pu to $0.835~$pu by applying the control $\uv[t] = \Bm^{-1} (\xv_0 - \widehat{\xv}[t])$. The attacker injects the attack vector on the sensor measurements of the pilot buses. The resulting pilot bus voltages and their estimates are plotted in Fig.~\ref{fig:IEEE39_voltage_state}. It can be observed that while the voltage estimates are close to the set-point, the actual pilot bus voltages deviate significantly from the set-point, thus showing the efficacy of the derived attacks.

\section{Conclusions}
\label{sec:Conclusion}
In this paper, we studied the performance of a CPCS with attack detection and reactive 
attack mitigation. We derived the optimal attack sequence that maximizes the state estimation error over the attack
time horizon using an MDP framework. Our results show that an arbitrarily constructed attack sequence will have little impact on the system since it will be detected, hence mitigated. The optimal attack sequence must be crafted to strike a balance between the stealthiness and the attack magnitude. Our results are useful for the system operator to assess the limit of attack impact and compare different attack detection and mitigation strategies. We also quantified the impact of FP and MD on the state estimation error, which 
helps select the right attack detection threshold depending on the accuracy of the attack mitigation signal.
We apply our results to the voltage control in a power system.

\bibliographystyle{IEEEtran}
\bibliography{bibliography}

\section*{Appendix A: Computation of the Cumulative State Estimation Error}

\subsection*{Appendix A - Part I}
We first state the result for the computation of $\mathbb{E}[\ev^{(i)}_d[t+1] ]$ and $\text{Var} \LB \ev^{(i)}_d[t+1]  \RB.$

The mean of $\ev^{(i)}_d[t+1]$ can be computed as
$$\mathbb{E}[\ev^{(i)}_d[t+1] ]  = \bar{\xv}^{(i)}  + \Sigmam_{\xv \yv}{\Sigmam^{-1}_{\yv \yv}}(\yv - \bar{\yv}),$$  
where,
$\bar{\xv}^{(i)} = \Am_K \mathbb{E}[\ev_c[t]] -   \Km (\av[t+1] - \mathbbm{1}_{\{i = 1 \}} \bar{\deltav}[t+1]) $, $\bar{\yv} = \Cm \Am \mathbb{E}[\ev_c[t]]  + \av[t+1] .$  The variance of $\ev^{(i)}_d[t+1]$ can be computed as
$$\text{Var} \LB \ev^{(i)}_d[t+1]  \RB = \Sigma^{(i)}_{\xv \xv} - \Sigmam_{\xv \yv}{\Sigmam^{-1}_{\yv \yv}} \Sigmam_{ \yv \xv},$$ 
where,
\begin{small}
 \begin{align*}
   &\Sigmam_{\yv \yv}  = \Cm \Am  \text{Var}(\ev_c[t]) \Am^T \Cm^T  + {\Cm} \Qm \Cm^T + \Rm,\\ 
   &\Sigmam_{\xv \yv}  = \Am_K  \text{Var}(\ev_c[t]) \Am^T \Cm^T + \Wm_K \Qm \Cm^T  - \Km \Rm^T, \\
   &\Sigmam_{\yv \xv}  = \Cm \Am  \text{Var}(\ev_c[t]) \Am_K^T   +\Cm \Qm  \Wm^T_K  - \Rm \Km^T,\\
   &\Sigmam^{(i)}_{\xv \xv} \!\!=\!\! \Am_K \text{Var}(\ev_c[t]) \Am^T_k \!\!+\!\! \Wm_K \Qm \Wm^T_K \!\!+\!\! \Km \Rm \Km^T \!\!+\!\! \mathbbm{1}_{ \{i \!=\! 1 \}} \Km \Bm \Km^T, 
 \end{align*}
 \end{small}
$\mathbbm{1}_{x} = 1$ if $x$ is true, or $0$ if $x$ is false, $\Bm = \mathbb{E} [(\deltav[t+1] - \mathbb{E}[\deltav[t+1]])(\deltav[t+1] - \mathbb{E}[\deltav[t+1])^T]$ and $f_{\yv}(y)$ is the Gaussian pdf with mean $\bar{\yv}$ and variance given by  $\Sigmam_{\yv \yv}.$
The integration limits $a_i$ and $b_i$ are  $[-\sqrt{\eta \Pm_r}, \sqrt{\eta \Pm_r}], \ \text{if} \ i = 0,$ and $(-\infty,-\sqrt{\eta \Pm_r}] \cup [\sqrt{\eta \Pm_r},\infty), \ \text{if} \ i = 1.$
The terms $\mathbb{E}[\ev_c[t]]$ and $\text{Var} (\ev_c[t])$ in the above can be in turn recursively
computed as 

\begin{small}
\begin{align*}
\mathbb{E}[\ev^{(i)}_c[t+1] ] = \frac{\int_{a_i}^{b_i} \mathbb{E}[\ev^{(i)}_d[t+1] ] f_{\yv}(y)}{\int_{a_i}^{b_i}f_{\yv}(y)};  \mathbb{E}[\ev_c[0] ] = 0,  i = 0,1,
\end{align*}
\end{small}
and 
$\text{Var} (\ev^{(i)}_c[t+1]) = \mathbb{E}[({\ev^{(i)}_c[t+1]})^2] - (\mathbb{E}[\ev^{(i)}_c[t+1]])^{2}; \  \text{Var} (\ev_c[0]) = \Pm_e.$
Finally the quantity  $f_{\iv_{[1:t]}} (\bv)$ can be computed recursively as 
$f_{\iv_{[1:t+1]}} (\bv) = \int_{a_i}^{b_i}  f_{\yv}(y), i = 0,1.$ 

\subsection*{Appendix A - Part II}

\begin{figure}
\begin{center}
  \def\svgwidth{0.6\columnwidth}
  \scriptsize
       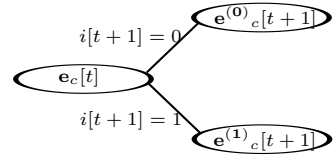
       \caption{Evolution of KF estimation error.}
       \label{fig:KF_Thm}
       \end{center}
       \vspace*{-1em}
\end{figure}
We now present the complete derivation of \eqref{eqn:mean_ec_final}. First, we present a sketch of the derivation. 
From Fig.~\ref{fig:KF_Thm}, note that $\ev^{(0)}_c[t+1]$ can be equivalently 
expressed as 
$$ \ev^{(0)}_c[t+1] = \ev_c[t+1] \big{|} \{ \iv_{[1:t]},\rv_c[t+1] \in  [-\sqrt{\eta \Pm_r}, \sqrt{\eta \Pm_r}] \}.$$
We first derive an expression for $\ev[t+1] \big{|} \{ \iv_{[1:t]}, \rv_c[t+1] = r \}$ where $r \in  [-\sqrt{\eta \Pm_r} , \sqrt{\eta \Pm_r}].$ Then, we evaluate $\mathbb{E} [(\ev^{(0)}_c[t+1])^2]$ by using $\mathbb{E} [X^2| a \leq Y \leq b] = \frac{\int_{a}^{b} \mathbb{E} [X^2| Y ] f_{Y}(y) }{\int_{a}^{b}f_{Y}(y)},$  where
$X = \ev^{(0)}_c[t+1], Y = \rv_c[t+1].$ The limits $a$ and $b$ correspond to $-\sqrt{\eta \Pm_r}$ and $\sqrt{\eta \Pm_r} $ respectively. We adopt a similar approach to evaluate $\mathbb{E} [(\ev^{(1)}_c[t+1])^2].$ We present the detailed derivation in the following.


For brevity we only present analysis in the case when $\rv_c[t+1] \in [-\sqrt{\eta \Pm_r}, \sqrt{\eta \Pm_r}].$  Recall the definitions of the terms  $\ev_c[t] =  \ev[t]  \big{|} \iv_{[1:t]},$
$\rv_c[t]  = \rv[t] \big{|} \iv_{[1:t-1]}.$ 
They evolve in time as 
\begin{multline}
 \ev_c [t+1]  =  \Am_K \ev_c[t] + \Wm_K\wv [t] \\- \Km (\av[t+1] -  i[t+1]   \deltav[t+1] )  - \Km \vv[t+1], \ t \geq 0,  \label{eqn:error_evol_cond} \\
 \end{multline}
 \begin{multline}
\rv_c[t+1]  = \Cm \Am \ev_c[t] + \Cm \wv[t] + \vv[t+1], \label{eqn:res_evol_cond}
\end{multline}
where the above are obtained by conditioning both sides of \eqref{eqn:error_evol} and the residual dynamics $\rv[t+1] = \Cm \Am \ev[t] + \Cm \wv[t] + \av[t+1] + \vv[t+1]$ on  $\iv_{[1:t]}$ and noting that $\wv[t]$ and $\vv[t]$ are independent of $\iv_{[1:t]}.$

For convenience, let us define 
\begin{multline}
\ev^{(0)}_d[t+1] \doteq \ev[t+1] \big{|} \{ \iv_{[1:t]}, \rv_c[t+1] = r \}  \\
 = \Am_K \ev_c[t]  + \Wm_K\wv [t]  - \Km \av[t+1] \\- \Km \vv[t+1] \big{|} \{ \rv_c[t+1] = r \},
\end{multline}
where in the above we have used the definition of $\ev_c[t]$ as well as the fact that $\wv [t],$ 
and $\vv[t+1]$ are independent of $\iv_{[1:t]}, \rv_c[t+1]$ and $i [t+1] = 0.$ 
Let us denote $\xv = \Am_K \ev_c[t]  + \Wm_K\wv [t] - \Km \av[t+1]   - \Km \vv[t+1]$ and $\yv = \rv_c[t+1].$ We note that $\xv$ and $\yv$ are Gaussian random variables and hence jointly Gaussian. Their mean 
and covariance matrix can be computed as
\begin{align*}
\mathbb{E}[\xv ] = \Am_K \mathbb{E}[\ev_c[t]] - \Km \av[t+1], \mathbb{E}[\yv ]\\
= \Cm \Am \mathbb{E}[\ev_c[t]] + \av[t+1]\ 
\ \text{and} \ \text{Cov} & (\Xm)    = \begin{bmatrix}
\Sigmam_{\xv \xv}  & \Sigmam_{\xv \yv} \\
\Sigmam_{\yv \xv } & \Sigmam_{\yv \yv}
\end{bmatrix},
\end{align*}
where them terms $\Sigmam_{\xv \xv}, \Sigmam_{\xv \yv}$ and $\Sigmam_{\yv \yv}$ are defined in Appendix~A, Part I (for $i = 0$). 

We now present in recursive method to compute $\mathbb{E}[\ev_c[t]]$ in the above expression.
First note based on our definitions, $ \ev^{(0)}_d[t+1]$ is the conditional random variable $\xv | \yv.$ Thus, the mean and variance of $ \ev^{(0)}_d[t+1]$ can be derived using multivariate normal distribution as 
\begin{align}
\mathbb{E} \LSB \ev^{(0)}_d[t+1] \RSB & = \mathbb{E}[\xv ] + \Sigmam_{\xv \yv}{\Sigmam_{\yv \yv}}^{-1}(\yv - \mathbb{E} [\yv]), \label{eqn:cond_expect} \\
\text{Var} \LB \ev^{(0)}_d[t+1]  \RB &= \Sigma_{\xv \xv} - \Sigmam_{\xv \yv}{\Sigmam_{\yv \yv}}^{-1} \Sigmam_{ \yv \xv} \label{eqn:cond_var}.
\end{align}
From \eqref{eqn:cond_expect}, we can compute the $\mathbb{E}[\ev^{(0)}_c[t+1] ]$ as
\begin{align}
\mathbb{E}[\ev^{(0)}_c[t+1] ] = \frac{\int_{-\sqrt{\eta \Pm_r} }^{\sqrt{\eta \Pm_r} } \mathbb{E}[\ev^{(0)}_d[t+1] ] f_{\yv}(\yv)}{\int_{-\sqrt{\eta \Pm_r} }^{\sqrt{\eta \Pm_r} }f_{\yv}(\yv)},
\end{align}
where we have used the fact that 
$\mathbb{E} [X| a \leq Y \leq b] = \frac{\int_{a}^{b} \mathbb{E} [X| Y ] f_{Y}(y) }{\int_{a}^{b}f_{Y}(y)}$. A similar method can be used to compute $\mathbb{E}[\ev^{(1)}_c[t+1] ].$ Then $\mathbb{E}[\ev_c[t+1] ]$ can be computed as 
$\mathbb{E}[\ev_c[t+1] ] = \mathbb{P} (i[t+1] = 0) \mathbb{E}[\ev^{(0)}_c[t+1] ] + \mathbb{P} (i[t+1] = 1) \mathbb{E}[\ev^{(1)}_c[t+1] ].$

The terms \eqref{eqn:cond_expect} and \eqref{eqn:cond_var} can also be used to 
compute $\mathbb{E} [(\ev^{(0)}_c [t+1])^2 ]$ as
\begin{align}
\mathbb{E}[(\ev^{(0)}_c[t+1])^2] & =  \frac{\int_{-\sqrt{\eta \Pm_r} }^{\sqrt{\eta \Pm_r}} \mathbb{E} \LSB  (\ev^{(0)}_d[t+1])^2  \RSB f_{\yv}(\yv) }{\int_{-\sqrt{\eta \Pm_r}}^{\sqrt{\eta \Pm_r} } f_{\yv}(\yv) } \label{eqn:here11},
\end{align}
where in we used the fact that $\mathbb{E} [X^2| a \leq Y \leq b] = \frac{\int_{a}^{b} \mathbb{E} [X^2| Y ] f_{Y}(y) }{\int_{a}^{b}f_{Y}(y)}.$ 
Note that integrand of \eqref{eqn:here11} is given by $$\mathbb{E} \LSB  (\ev^{(0)}_d[t+1])^2  \RSB = \LB \mathbb{E} \LSB  \ev^{(0)}_d[t+1]  \RSB \RB^2 + \text{Var} \LB \ev^{(0)}_d[t+1] \RB,$$
where the RHS terms are in \eqref{eqn:cond_expect} and \eqref{eqn:cond_var} respectively. 
We remark that the case $\rv_c[t+1] \notin [-\sqrt{\eta \Pm_r} , \sqrt{\eta \Pm_r} ]$
can be similarly analyzed to derive $\mathbb{E}[(\ev^{(1)}_c[t+1])^2].$
We skip the derivation here and The final result is stated in Appendix~A, Part I.

Finally, the quantity $f_{\iv_{[1:t]}} (\bv)$ can be computed 
by noting that $\mathbb{P} (i[t+1] = 0 | \iv_{[1:t]}) = \mathbb{P} (\rv_c[t+1] \in [-\sqrt{\eta \Pm_r} , \sqrt{\eta \Pm_r}] ) $ and $\mathbb{P} (i[t+1] = 1 | \iv_{[1:t]}) = \mathbb{P} (\rv_c[t+1] \notin [-\sqrt{\eta \Pm_r} , \sqrt{\eta \Pm_r}]).$ The quantities $\mathbb{P} (\rv_c[t+1] \in [-\sqrt{\eta \Pm_r} , \sqrt{\eta \Pm_r}])$ and $\mathbb{P} (\rv_c[t+1] \notin [-\sqrt{\eta \Pm_r}, \sqrt{\eta \Pm_r}])$ can be computed by noting that $\rv_c[t+1]$ is a Gaussian distributed random variable
from with mean $\mathbb{E}[\yv]$ and variance $\Sigmam_{\yv \yv}$ (from \eqref{eqn:res_evol_cond}).

\begin{IEEEbiography}
[{\includegraphics[width=1in,height=1.25in,clip,keepaspectratio]{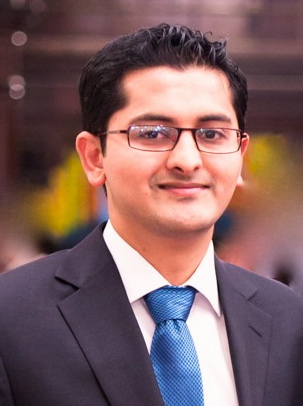}}]
{Subhash Lakshminarayana} (S'07-M'12) is an assistant professor at the School of Engineering, University of Warwick, UK. Previously, he worked as a researcher at the Advanced Digital Sciences Center (ADSC) in Singapore between 2015-2018, a joint post-doctoral researcher at Princeton University and the Singapore University of Technology and Design (SUTD) between 2013-2015. He received his Ph.D. from the
Alcatel Lucent Chair on Flexible Radio
and the Department of Telecommunications at SUPELEC, France in 2012, M.S. degree in Electrical
and Computer Engineering from The Ohio State
University in 2009 and B.S. from Bangalore University, India. 

His research interests include cyber-physical system security (power grids and urban transportation) and wireless communications. His works have been selected among the Best conference papers on integration of renewable \& intermittent resources at the IEEE PESGM - 2015 conference, and the ``Best 50 papers" of IEEE Globecom 2014 conference.


\end{IEEEbiography}

\begin{IEEEbiography}
[{\includegraphics[width=1in,height=1.25in,clip,keepaspectratio]{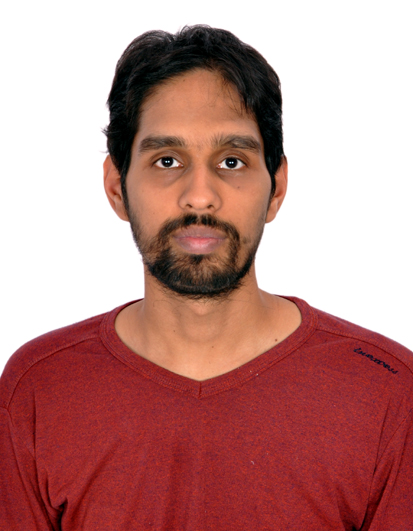}}]
{Jabir Shabbir Karachiwala} obtained Masters degree from the School of Computing
in National University of Singapore in 2015. Between 2016-2018, he worked as research engineer and software engineer
at Advanced Digital Science Centre and Singapore MIT Alliance Of Research And Technology 
in Singapore on topics pertaining to transportation and cybersecurity. Currently he is working as Member of technical Staff
at Oracle India Pvt Ltd.

\end{IEEEbiography}


\begin{IEEEbiography}[{\includegraphics[width=1in,height=1.25in,clip,keepaspectratio]{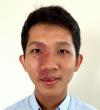}}]
{Teo Zhan Teng} was a research engineer at the Advanced Digital Sciences Center, Singapore from 2015-2016. Currently, he is working with GovTech, Singapore. 

\end{IEEEbiography}

\begin{IEEEbiography}[{\includegraphics[width=1in,height=1.25in,clip,keepaspectratio]{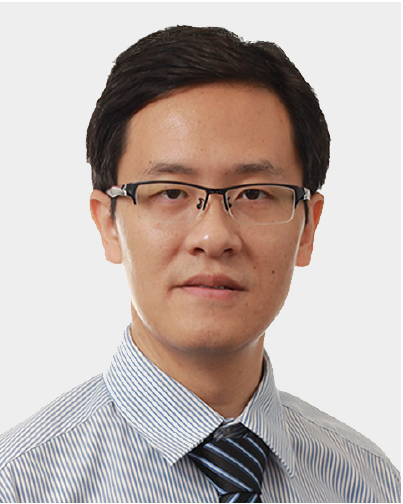}}]
{Rui Tan} (M'08-SM'18) is an Assistant Professor at School of Computer Science and Engineering, Nanyang Technological University, Singapore. Previously, he was a Research Scientist (2012-2015) and a Senior Research Scientist (2015) at Advanced Digital Sciences Center, a Singapore-based research center of University of Illinois at Urbana-Champaign (UIUC), a Principle Research Affiliate (2012-2015) at Coordinated Science Lab of UIUC, and a postdoctoral Research Associate (2010-2012) at Michigan State University. He received the Ph.D. (2010) degree in computer science from City University of Hong Kong, the B.S. (2004) and M.S. (2007) degrees from Shanghai Jiao Tong University. His research interests include cyberphysical systems, sensor networks, and pervasive computing systems. He received the Best Paper Awards from IPSN’17, CPSR-SG’17, Best Paper Runner-Ups from IEEE PerCom’13 and IPSN’14.
\end{IEEEbiography}

\begin{IEEEbiography}[{\includegraphics[width=1in,height=1.25in,clip,keepaspectratio]{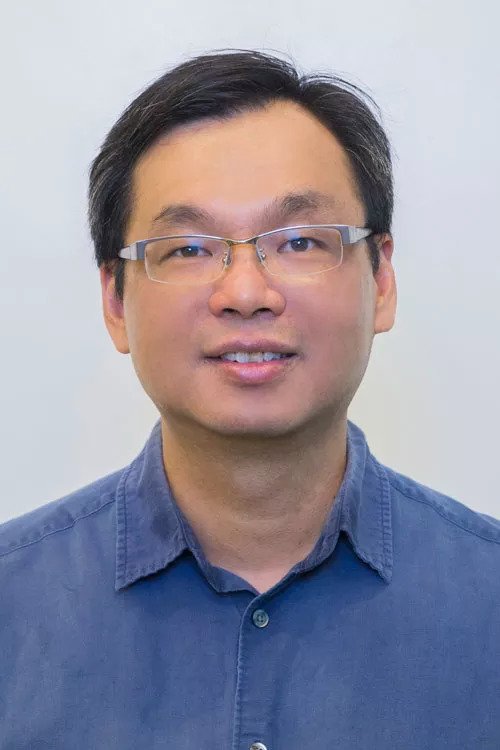}}]
{David K.Y. Yau} received the B.Sc. from the Chinese University of Hong Kong, and M.S. and Ph.D. from the University of Texas at Austin, all in computer science. He has been Professor at Singapore University of Technology and Design since 2013. Since 2010, he has been Distinguished Scientist at the Advanced Digital Sciences Centre, Singapore. He was Associate Professor of Computer Science at Purdue University (West Lafayette). He received an NSF CAREER award. He won Best Paper award in 2017 ACM/IEEE IPSN and 2010 IEEE MFI. His papers in 2008 IEEE MASS, 2013 IEEE PerCom, 2013 IEEE CPSNA, and 2013 ACM BuildSys were Best Paper finalists. His research interests include cyber-physical system and network security/privacy, wireless sensor networks, smart grid IT, and quality of service. He serves as Associate Editor of IEEE Trans. Network Science and Engineering and ACM Trans. Sensor Networks. He was Associate Editor of IEEE Trans. Smart Grid, Special Section on Smart Grid CyberPhysical Security (2017), IEEE/ACM Trans. Networking (2004-09), and Springer Networking Science (2012-2013); Vice General Chair (2006), TPC co-Chair (2007), and TPC Area Chair (2011) of IEEE ICNP; TPC co-Chair (2006) and Steering Committee member (2007-09) of IEEE IWQoS; TPC Track co-Chair of 2012 IEEE ICDCS; and Organizing Committee member of 2014 IEEE SECON. He is a senior member of IEEE.

\end{IEEEbiography}

\end{document}